\documentclass[prd,twocolumn,preprintnumbers,nofootinbib]{revtex4}
\usepackage{graphicx}
\usepackage{color}
\usepackage{float}
\newcommand{\rthis}[1]{\textcolor{black}{#1}}
\usepackage{amsfonts,amsmath,amssymb}
\usepackage[plainpages=false, colorlinks=true, anchorcolor=blue, linkcolor=blue, citecolor=blue, bookmarks=false]{hyperref}
\usepackage{natbib}
\usepackage{enumitem}

\pdfoutput=1
\begin{document}
%\markboth{Authors' Names}
%\newcommand{\rthis}[1]{\textcolor{black}{#1}}
\newcommand{\apjl}{Astrophys. J. Lett.}
\newcommand{\apjs}{Astrophys. J. Suppl. Ser.}
\newcommand{\aap}{Astron. \& Astrophys.}
\newcommand{\aj}{Astron. J.}
\newcommand{\araa}{Ann. Rev. Astron. Astrophys. } %ARA$\&$A}
\newcommand{\mnras}{Mon. Not. R. Astron. Soc.}
\newcommand{\ssr}{Space Science Revs.}
\newcommand{\apss}{Astrophysics \& Space Sciences}
\newcommand{\jcap}{JCAP}
\newcommand{\pasj}{PASJ}
\newcommand{\pasp}{PASP}
\newcommand{\pasa}{Pub. Astro. Soc. Aust.}
\newcommand{\physrep}{Phys. Rep.}
\title{Yet another test  of Radial Acceleration Relation for galaxy clusters}
\author{S. \surname{Pradyumna}$^1$}
 \altaffiliation{E-mail:ep18btech11015@iith.ac.in}
\author{Sajal \surname{Gupta}$^2$ }
\altaffiliation{E-mail:sajalgupta.04@gmail.com}
\author{Sowmya \surname{Seeram}$^3$}  \altaffiliation{E-mail:ep17btech11018@iith.ac.in}
\author{Shantanu  \surname{Desai}$^4$} \altaffiliation{E-mail: shntn05@gmail.com}
\affiliation{$^{1,3,4}$Department of Physics, Indian Institute of Technology, Hyderabad, Telangana-502285, India}
\affiliation{$^{2}$Department of Physical Sciences, IISER-Kolkata, Mohanpur, West Bengal-741246, India}

\begin{abstract}
We carry out a test of the radial acceleration relation  (RAR) for galaxy clusters from two different catalogs compiled in literature, as an independent  cross-check of  two recent analyses, which reached opposite conclusions. The  datasets we considered include a Chandra  sample  of 12 clusters and the X-COP sample of 12 clusters.  \rthis{For both the samples,  we find that  the residual scatter is small (0.11-0.14 dex), although the best-fit values for the Chandra sample have large error bars. Therefore, we argue that at least one of these  cluster samples (X-COP) obeys the radial acceleration relation. However, since  the best-fit parameters are discrepant with each other as well as  the  previous estimates, we argue  that the RAR is not universal.} For both the catalogs, the acceleration scale, which  we obtain is about an order of magnitude larger than  that obtained for galaxies, and is agreement with  both the  recent estimates.
\end{abstract}

\maketitle

\section{Introduction}
Recently, two independent groups~\cite{Chan20,Tian} carried out  a test of the correlation between baryonic and total acceleration for two different cluster samples. This was motivated by the discovery of a strong deterministic relation, between the baryonic  ($a_{bar}$)   and total acceleration ($a_{tot}$) using 2693 data points from 153 rotationally supported spiral galaxies of different sizes and morphology, with a  small scatter of \rthis{0.13} dex~\cite{Mcgaugh16,Lelli}. This scatter is dominated by uncertainties in galaxy distance, disk inclination and the stellar mass-to-light ratio. Once these are marginalized over, the residuals around the fits have negligible scatter (of about 0.057 dex)~\cite{Li18}.  \rthis{Note however that this aforementioned scatter  was obtained  by assuming that every galaxy obeys the mean RAR, followed by doing a separate fit for each of the 175 galaxies, followed by marginalizing over the aforementioned uncertainties.}
A similar relation has also been recently found for elliptical galaxies~\citep{Sheth}.
This relation has been dubbed  as  the radial acceleration relation (RAR) and can be written as follows:
\begin{equation}
a_{tot} = \frac{a_{bar}}{1-e^{-\sqrt{a_{bar}/a_0}}},
\label{eq:RAR}
\end{equation}
where $a_0 \sim 1.2 \times 10^{-10} m/s^2$~\cite{Mcgaugh16}. 
This equation is closely related to the Mass Discrepancy-Acceleration relation~\cite{Mcgaugh04}, and is  a trivial consequence of the MOND paradigm~\cite{Famaey12,Milgrom16}, and also subsumes the baryonic Tully-Fisher relation~\cite{Lelli17,Wheeler}.  We note that a few  groups have  have  questioned the existence of a fundamental acceleration scale in the same data~\cite{Rodrigues,Zasov,Marra,Zhu20}. Although, some of these criticisms have been countered~\cite{McGaughNature,Kroupa}, the debate is still on-going on some of these issues~\cite{delpopolo18,Rodrigues20}.
There are mixed claims in the literature on  whether this  relation can be easily postdicted using the standard  $\Lambda$CDM  paradigm~\cite{Desmond,Navarro,Lu,Keller,Garaldi,Ludlow,Stone}, with the same negligible scatter as observed for real data. This relation can also be reproduced using alternative models such as Self-Interacting Dark matter~\cite{Ren}, superfluid dark matter~\cite{Khoury}, Moffat's MOG~\cite{Green}, conformal gravity~\cite{Mannheim},  but is  in tension with Verlinde's entropic gravity~\cite{Schombert}.  

Motivated by these considerations, two independent groups~\cite{Chan20,Tian}  carried out a systematic observational test of the RAR using galaxy clusters. 
Galaxy clusters are the most massive collapsed objects in the universe, and have proved to be a  wonderful laboratory for a wide range of topics from cosmology, tests of modified gravity,  fundamental Physics (such as constraining neutrino and graviton mass) to galaxy evolution~\cite{Voit,Vikhlininrev,Allen,Desai18}. 
However, it has  been known since a long time, that MOND does not work well for both relaxed as well as merging galaxy clusters~\cite{White88,Gerbal,Aguirre,Sanders,Silk,Angus,Natarajan}, and is also at odds with shape of the matter power spectrum~\cite{Dodelson}. Many relativistic theories of MOND are also in tension due to the  coincident gravitational wave and electro-magnetic observations from GW170817~\citep{Woodard}.

Nevertheless, there was  no detailed characterization of the RAR for clusters, prior to the two works~\cite{Chan20,Tian}. Despite the failures of MOND on cluster scales, a detailed characterization of the RAR for a large ensemble of galaxy clusters would be an acid test for some of the myriad models used to explain the same relation for galaxies. However, there are also no clear cut predictions for whether the RAR should be satisfied or not for galaxy clusters, except for a few works~\cite{Famaey18,Navarro}, which predict that this relation should not be seen for galaxy clusters. We briefly discuss the results from these two observational tests of  RAR for clusters. 

Chan and Del Popolo~\cite{Chan20} (CDP20, hereafter)
considered 52 non-cool-core clusters in hydrostatic equilibrium from the HIFLUGCS X-ray sample, based on ROSAT and ASCA observations~\cite{Chen},  with core-radius ($r_c$) greater than 100~kpc. 
The density of the hot gas was modeled using a single-$\beta$ profile~\cite{betamodel} and a constant temperature profile was assumed.
Then assuming hydrostatic equilibrium, the total mass was calculated
assuming spherical symmetry~\cite{Allen}. From the total mass, the total acceleration was estimated assuming Newtonian gravity~\cite{Chan20}. Similarly, the baryonic acceleration was obtained from the baryonic mass (estimated by integrating the gas density profile). We note that the stellar mass was not included in their baryonic mass budget. CDP20 then checked for a correlation between the baryonic and total acceleration at four different positions ($r_c$, $2r_c$, $3r_c$, $r_{500}$) for their sample. They found that  the best-fit value for $a_0$ was $9.5 \times 10^{-10} m/s^2$, about ten  times higher than the same found for spiral galaxies. Furthermore, their estimated \rthis{residual scatter which they obtained when  compared to the RAR}  is about 0.18 dex,  and much higher than that for galaxies.

Around the same time, Tian et al~\cite{Tian} (T20, hereafter) tested this relation for a sample of 20 high-mass clusters from the CLASH survey~\cite{Postman}. These clusters have a combination of strong-lensing, weak lensing shear, and magnification data using HST and Suprime-Cam observations~\cite{Umetsu16}; and gas temperatures and density profiles  from both Chandra and XMM observations~\cite{Donahue14}. The total mass was modeled as a sum of four distinct components: the total dark matter, X-ray emitting gas, non-BCG stellar mass and BCG stellar mass. The baryonic mass was obtained from the sum of the last three components. The dark matter mass was obtained from the NFW~\cite{NFW} density profile and mass-concentration relations from ~\cite{Umetsu16}. The X-ray gas mass was obtained from ~\cite{Donahue14}. The non-BCG stellar mass was obtained from the stellar mass fraction derived using the SPT sample~\cite{Chiu18}. The BCG stellar mass was estimated by assuming an Hernquist profile~\cite{Hernquist} for the stellar mass distribution and assuming the the stellar mass estimates from Ref.~\cite{Cooke}. They looked for RAR  by transforming $a_{tot}$ and $a_{bar}$ into log-log space, thereby converting it to a linear regression problem. They obtained an extremely tight scaling relation between the total and baryonic acceleration, with an intrinsic scatter of about 14\%.   This is comparable to the scatter obtained for the SPARC sample~\cite{Mcgaugh16}. However, the best-fit value for $a_0$  was $2 \times 10^{-9} m/ s^2$, about 20 times higher. They also explained the observed scatter using a semi-analytic galaxy formation model within $\Lambda$CDM~\cite{Olamaie}. Therefore, although both of these works get about the same value for the acceleration scale, they disagree on whether the RAR exists for galaxy clusters with negligible scatter. We note that most recently, this group has also found evidence for a tight Baryonic Faber-Jackson relation for the HIFLUGCS cluster sample~\cite{Ko20}

Given the contradictory findings between the two groups,  we carry out a similar test of the RAR for  two different cluster samples. The first dataset we analyze is a sample of 12 galaxy clusters with pointed Chandra and archival ROSAT observations which were carefully studied  by ~\citet{Vikhlinin05,Vikhlinin06}.  This dataset has been used for testing a plethora of modified gravity theories~\citep{Rahvar,Hodson17,Khoury17,Ng,Edmonds,Bernal}. We have also previously used this sample to constrain the graviton mass, test of constant halo surface density, and  GR corrections to hydrostatic masses~\cite{Gupta1,Gupta2,Gopika}. 

We  now provide a brief clarification of what we mean by  a test of RAR for clusters. Although the moniker ``RAR'' (for spiral galaxies) usually refers to the empirical relation in Eq.~\ref{eq:RAR} with the same  value of $a_0$ as found in ~\cite{Mcgaugh16}, one can get very similar deterministic relations between $a_{tot}$ and $a_{bar}$ for other proposed MOND interpolating functions~\cite{Famaey12}. A test of these relations for spiral galaxies would also yield a very tight scatter, although the difference between these different functions could be discerned using the Halo Acceleration relation~\citep{Tian19}. Similarly,  it is possible that  for  galaxy clusters, Eq.~\ref{eq:RAR} could fit the data, albeit with a higher value for  $a_0$, similar to the value for $a_0$ found in the first ever test of MOND with clusters~\cite{White88}. To carry out a generic model-independent test, we note that Eq.~\ref{eq:RAR} can be re-written as  a linear  regression equation between the logarithm of the total and baryonic acceleration (cf. Eq.~\ref{lineareq}). This is the same fit done in T20. Therefore, for the rest of this manuscript by ``test  of RAR'', implies a  test  of whether there exists  \rthis{a tight empirical  scaling relation between the baryonic and total accelerations in logarithmic space},  throughout the galaxy cluster. \rthis{We shall also characterize the scatter in the residuals between the two accelerations, as well as  determine the intrinsic scatter in the scaling relations. A separate question is whether this relation is universal, or in other words the different cluster samples yield the same best-fit values. We shall also test for the universality of this relation for our cluster samples.} 

\section{Basic considerations}

We review the  Physics,  common to all galaxy cluster samples, needed to estimate the total acceleration as well as the baryonic acceleration.

In the first step, we calculate the total mass from equation of hydrostatic equilibrium, which can be applied if the clusters are relaxed. The systematic uncertainty in this assumption is about 15-20\%~\cite{Biffi16}.
If the parametric forms for the gas temperature and gas density profiles are given, the total acceleration at given radius $r$ from the cluster center  can be derived from  positing an ideal gas equation of state~\cite{Allen}:
\begin{equation}
a_{tot} (r) = -\frac{k_bT }{r\mu m_p}\left(\frac{d\ln\rho_{gas}}{d\ln r}+ \frac{d\ln T}{d\ln r}\right),
\label{eq:acc}
\end{equation}
where $\rho_{gas}$ denotes the gas mass density, $T$ is the cluster temperature, $m_p$ is the mass of the proton,  and $\mu$ is the mean molecular weight of the cluster in a.m.u. and is approximately equal  to 0.6~\cite{Vikhlinin05,Rahvar}.  Alternately, if the total total mass  ($M_{tot}$) is known, then the total acceleration is given by:
\begin{equation}
a_{tot}=\frac{G M_{tot}}{r^2}
\label{eq:atotfrommass}
\end{equation}
%\begin{equation}
%a_{tot} = \frac{1}{\mu m_p n_e}\frac{dP}{dr}  
%\label{eq:accpress}
%\end{equation}

In order to obtain the baryonic acceleration, we need to estimate the total mass in baryons, which consists of hot diffuse gas in the intra-cluster medium and stars. The gas mass is obtained by integrating the gas density profile:
\begin{equation}
M_{gas} = \int 4 \pi r^2 \rho_{gas}(r) dr  
\label{eq:mgas}
\end{equation}
We note  that the aforementioned estimates of gas mass and the total mass assume spherical symmetry. Errors due to this assumption can be upto 5\%~\cite{Gopika}. 
There are multiple methods to obtain the star mass in galaxy clusters~\cite{ytlin,Chiu18}. We use the method in ~\cite{ytlin}. This entails calculating   the total mass ($M_{500}$) within a radius with an overdensity of 500, called $r_{500}$. We then use the following empirical relation to obtain the star mass
at $r=r_{500}$~\cite{ytlin}.
\begin{equation}
    \frac{M_{star} (r=r_{500})}{10^{12} M_{\odot}} = 1.8 \left(\frac{M_{500}}{10^{14} M_{\odot}}\right)^{0.71}
\label{eq:mstar}
\end{equation}
%Please no changes in the manuscript now as its under review
This equation yields the star mass at $r=r_{500}$. The star mass at any other radius can be obtained by assuming an isothermal profile~\cite{Rahvar}. We note that since the stellar mass contribution is negligible, errors due to stellar mass estimate should not dominate the total budget and are neglected.

The baryonic acceleration is then given by
\begin{equation}
a_{bar} = \frac{G (M_{gas}+M_{stars})}{r^2}
\label{eq:abar}
\end{equation}

\section{Chandra X-ray cluster sample}
\label{sec:chandra}
Vikhlinin et al~\cite{Vikhlinin05,Vikhlinin06} provided gas density and temperature profiles for 13 different clusters, using measurements  from the pointed or archival observations with  the Chandra  and 
ROSAT X-ray satellites. For our analysis we use 12 of these clusters, viz.  A133, A262, A383, A478, A907, A1413, A1795, A1991, A2029, A2390, MKW4, RXJ1159+55531, since errors for one of them were not available. The redshifts of these clusters range approximately upto $z=0.2$. These measurements extended up to very large radii  of about $r_{500}$ for some of the clusters.  The typical exposure times ranged from 30-130 ksecs. The temperatures span the range between  1 and 10 keV and masses from $(0.5-10) \times 10^{14} M_{\odot}$.  More details of the observations, data reduction, and estimation of temperature and density profiles have been discussed in \citet{Vikhlinin06}.
This dataset has been used for testing a plethora of modified gravity theories~\citep{Rahvar,Hodson17,Khoury17,Ng,Edmonds,Bernal,Gupta1}. 
%We have previously used this sample to constrain the graviton mass, assess effect of relativistic corrections and test of constancy of dark matter halo surface density~\cite{Gupta1,Gupta2,Gopika}.
We note that although the first  test of RAR for this sample  was done in ~\cite{Ng}, no error analysis in the observed accelerations was taken into account in that analysis. They considered 250 radii points starting at 10 pc and extending to 2.5 kpc in increments of 10 pc (D. Edmonds, private communication). Also the stellar mass was not included in their estimation of baryonic mass. Based on a visual inspection of the scatter plot between $a_{tot}$ and $a_{bar}$, ~\citet{Ng} asserted that there is a large scatter about the RAR. However, since no error bars were included in that work, no quantitative estimate of the intrinsic scatter was done.

The gas density ($\rho_g$) model used for this sample is  the double-beta profile~\cite{Vikhlinin06} and is equal to  $1.624 m_p\sqrt{n_p(r)n_e(r)}$, where $m_p$ is the proton mass and 
the product of electron and proton densities ($n_e n_p$) is given by the double-$\beta$ profile: 
\begin{widetext}
\begin{equation}
n_e (r)n_p(r) = \frac{(r/r_c)^{-\alpha'}}{(1+r^2/r_c^2)^{3\beta -\alpha' /2}}\frac{n_0^2}{(1+r^\gamma/r_s^\gamma)^{\epsilon/\gamma}} + \frac{n_{02}^2}{(1+r^2/r_{c'}^{2})^{3\beta'}}.
\label{eq:gasdensity}
\end{equation}
\end{widetext}
The explanation and values for  each of the terms in Eq.~\ref{eq:gasdensity}  can be found in the original works by Vikhlinin~\cite{Vikhlinin09}. The temperature profile is given by:
\begin{equation}
T(r) = T_0 \frac{(x_0+T_{min}/T_0)}{x_0+1}\frac{(r/r_t)^{-a'}}{\left[1+(r/r_t)^b\right]^{c'/b}},
\label{eq:temp}
\end{equation}
where $x_0=\left(\frac{r}{r_{cool}}\right)^{a_{cool}}$. The  physical explanation of the eight free  parameters   and their  values for the clusters can be found in    ~\cite{Vikhlinin06,Rahvar}. The observed values of $T(r)$ along with their error bars  at various points from the cluster center were provided for the 12 clusters by A. Vikhlinin (private communication). 

We now plug in the values for $\rho(r)$ and $T(r)$ from Eqns.~\ref{eq:gasdensity} and \ref{eq:temp} respectively in Eq.~\ref{eq:acc} to calculate the total acceleration $a(r)$. The errors were obtained by propagating the $T(r)$ errors provided in V06. We then estimate the baryonic mass from the  sum of gas mass (obtained from Eq.~\ref{eq:mgas}) and star mass (obtained from  Eq.~\ref{eq:mstar}).

Therefore, from Eqns.~\ref{eq:acc} and \ref{eq:abar}, $a_{tot}$ and $a_{bar}$  can be calculated at any distance $r$ from the cluster center. As mentioned earlier, CDP20 have done this test at different radii for the various clusters at integer multiples of $r_c$, whereas  T20 have stacked data from the CLASH cluster sample (of about 10-18 clusters) at four different radii, with a total of 64 datapoints. 

Here, we do a joint fit to the RAR using the data for $a_{tot}$ and $a_{bar}$ using data at three different radii: 200 kpc, 500 kpc, and 1000 kpc . Similar to T20, we do a fit in log-acceleration space by doing a linear regression to the relation: 
\begin{equation}
y=m x + b
\label{lineareq}
\end{equation}
where $y=\ln a_{tot}$ and $x=\ln a_{bar}$. 
We then  maximize a log-likelihood  given by :
\begin{eqnarray}
-2\ln L &=& \large{\sum_i} \ln 2\pi\sigma_i^2 + \large{\sum_i} \frac{[y_i-(mx_i+b)]^2}{\sigma_i^2}
\label{eq:eq13}  \\
\sigma_i^2 &=& \sigma_{y_i}^2+m^2\sigma_{x_i}^2+\sigma_{int}^2
\label{eq:error}
\end{eqnarray}

Here, $\sigma_{y_i}$ and $\sigma_{x_i}$ denote the errors in $\ln a_{tot}$  and $\ln a_{bar}$ respectively; $\sigma_{int}$ is the intrinsic scatter for our linear relation between the logarithms of the accelerations, which is a free parameter. 
We used the {\tt emcee} MCMC sampler~\cite{emcee} to maximize Eq.~\ref{eq:eq13}. Our best-fit values for $m$
and $b$ are shown in Table~\ref{tab:summary}, along with the corresponding fits from CDP20 and T20. The 68\%, 90\% and 99\% marginalized confidence intervals for $m$, $b$, and $\sigma_{int}$ can be found in Fig.~\ref{fig:chandra_rar_corner}. Although the intrinsic scatter is small, the error on the intercept is quite large. As we can see, the 1-D marginalized credible interval for $\ln(\sigma_{int})$ remains flat. We also verified that alternate samplers such as {\tt Nestle}\footnote{\url{http://kylebarbary.com/nestle/}}  and  analytical optimizaion routines return the same best-fit as {\tt emcee} to  within $1\sigma$ error bars, as well as similar credible intervals. As the contribution to the total scatter from the best-fit value of $\sigma_{int}=2\times 10^{-4}$ is negligible, smaller values for $\sigma_{int}$ will not change the likelihood.  That explains the flatness of the marginalized distribution of $\ln(\sigma_{int})$. Our slope and intercept are  discrepant  compared to T20 at about $2.3\sigma$ and $2.2\sigma$ respectively.

\rthis{In order to compare the tightness of our RAR relation for this sample, we also calculate the residual scatter in dex to directly compare with the results in CDP20 and T20 (and also~\cite{Mcgaugh16,Li18}), who used a similar procedure to characterize the scatter. The residuals are given by $\log_{10} (a_{tot}) - \log_{10} (a_{exp})$, where $a_{exp}$ is the expected total acceleration obtained from the best-fit RAR. These binned residuals (using four bins) are then fitted to a Gaussian and the standard deviation of this Gaussian distribution gives the residual scatter in dex. When we fit these residuals for the Chandra sample to a Gaussian distribution, we get $\sigma=0.13$ dex. However, we  find that the residuals are asymmetric with longer tails at negative values for the residuals. Therefore, instead of the Gaussian $\sigma$, we use  the scale parameter obtained using the normalized median absolute deviation, $\sigma_{MAD}$, (computed using the {\tt astropy}~\cite{astropy} library) to parametrize the residual scatter. This value is given by  $\sigma_{MAD}=0.14$. Therefore, the residual scatter for the Chandra sample  is equal to 0.14 dex.}

\begin{figure}
\centering
\includegraphics[width=0.5\textwidth]{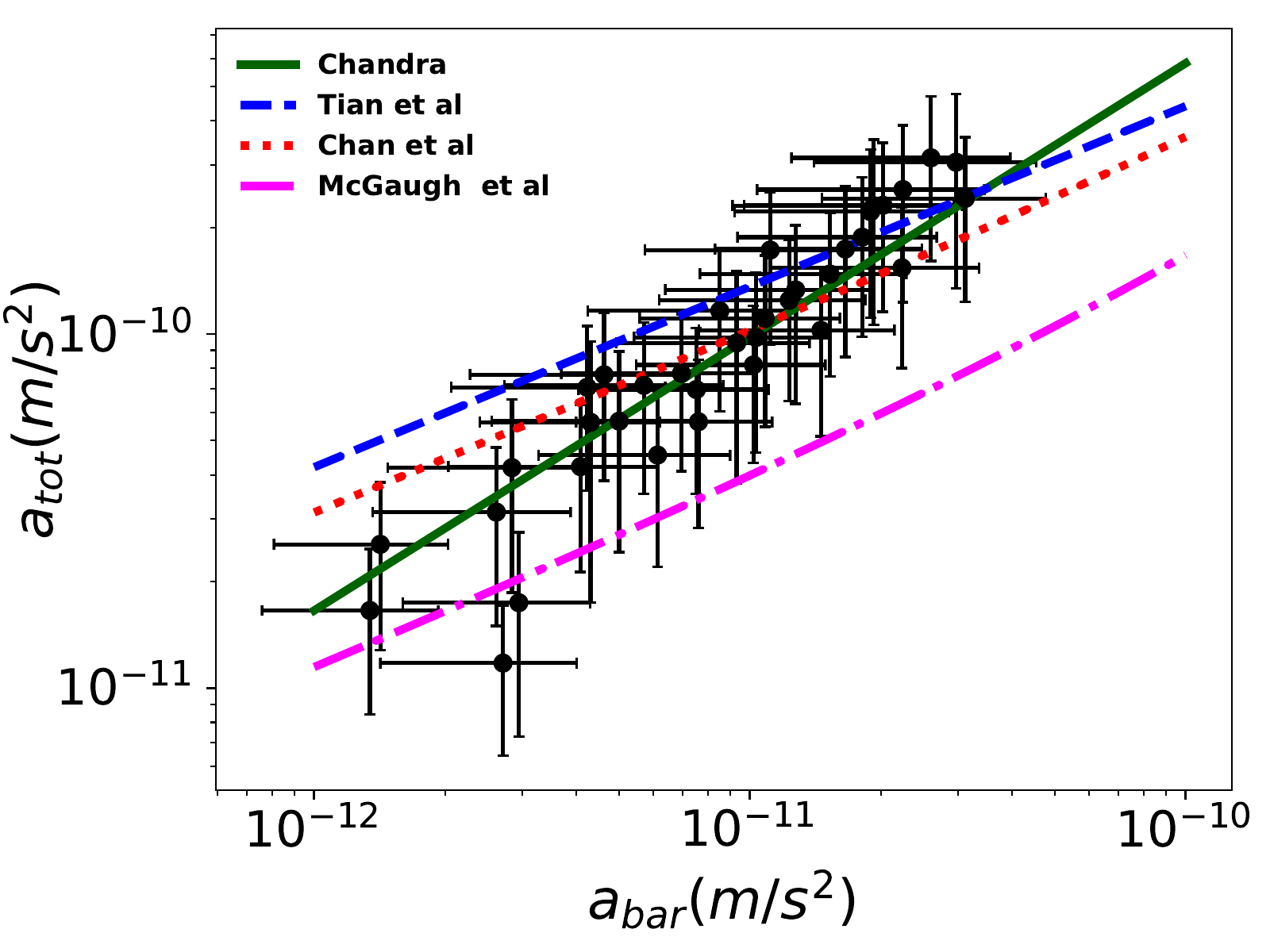}
\caption{A test of the radial acceleration relation for 12 clusters from the Chandra sample (black dots) ~\cite{Vikhlinin06}
using data stacked at 200 kpc, 500 kpc, and  1000 kpc.  
We also overlay our best-fit along with that from T20, CDP20 as well as the best-fit  RAR deduced from the SPARC sample~\cite{Mcgaugh16}. We see that the SPARC-based RAR does not agree with  the Chandra data.}   
\label{fig1}
\end{figure}

\begin{figure*}[h]
    \centering
    \includegraphics[width=15cm, height=13cm]{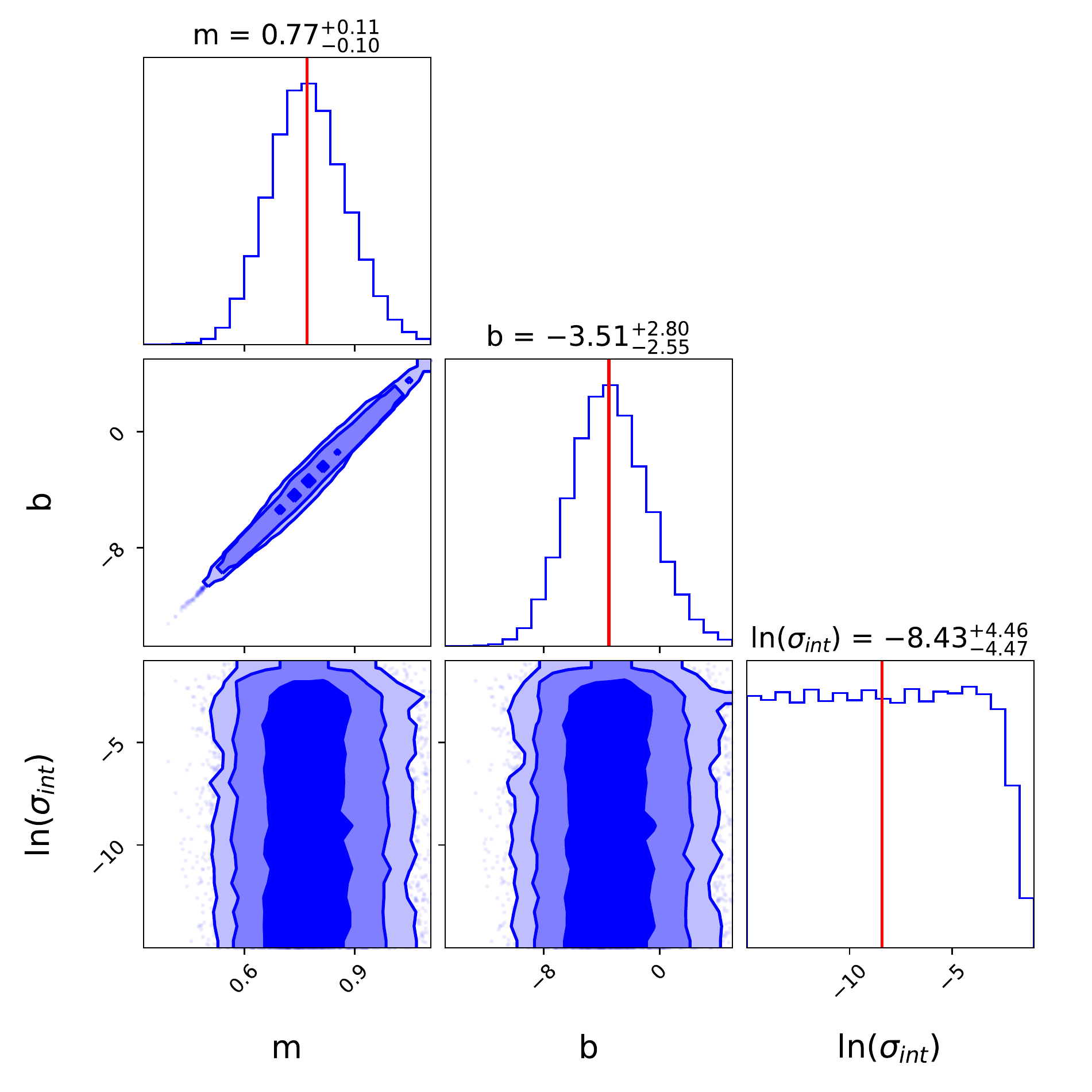}
    \caption{Plot showing 68\%, 90\% and 99\% marginalized credible intervals for the slope, intercept and  (natural log) intrinsic scatter for the linear relation between $\ln a_{tot}$ and $\ln a_{bar}$ for the Chandra cluster sample.}
    \label{fig:chandra_rar_corner}
\end{figure*}

We overlay our best-fit values along with the same from T20 and CDP20, as well as the RAR found from the  SPARC sample~\cite{Mcgaugh16} in Fig.~\ref{fig1}. Because of the large errors, both  T20 and CDP20 fits pass through the data points. However,  despite the large errors, the RAR obtained from SPARC sample is below most of the  data points.

To determine the  acceleration scale $a_0$, similar to T20, we fix the slope in the regression relation in Eq.~\ref{lineareq} to be equal to 0.5. This way the RAR relation (Eq.~\ref{eq:RAR}) in the low acceleration limit becomes $a_{dyn}=\sqrt{a_0 a_{bar}}$. From this relation it is trivial to derive the Baryonic Tully-Fisher or Faber-Jackson relations~\cite{Famaey12}.
The resulting value of $a_0$ is equal to $(9.26 \pm 1.66) \times 10^{-10} m/s^2$. Therefore, similar to T20 and CDP20, the value of $a_0$ is about an order of magnitude larger than obtained using galaxies~\cite{Mcgaugh16}. Our estimated  value  for $a_0$ agrees within our $1\sigma$ errors with  the estimate in CDP20.

\section{X-COP sample}
\label{sec:xcop}
X-COP (\textit{XMM-Newton} Cluster Outskirts Project)  (PI: D.Eckert) is a very large program on XMM-Newton dedicated to studying
X-ray emission in cluster outskirts.
The  X-COP project targeted a sample of massive 
12 galaxy clusters ($M_{500}>3\times 10^{14} M_{\odot}$) in the redshift range of $0.04-0.1$ selected from the Planck all-sky Sunyaev-Zel'dovich (SZ) catalog~\cite{planck_2014}. These selected clusters had  a   signal-to-noise ratio cutoff of 12.  The detailed  cluster selection criterion is outlined in ~\citet{xcop_thermo}. The final clusters included in this project are -  A85, A644, A1644, A1795, A2029, A2142, A2255, A2319, A3158, A3266, RXC1825, and ZW1215.~\cite{XCOPmain}.
The main goal of this project is to reconstruct the ICM properties upto $R_{200}$.
 More details on the XCOP sample and associated science results using this sample can be found in \cite{XCOP1,xcop_thermo,xcop_thermoa2319,xcop_He,Ghizzardi}. This cluster sample has also been  used to test modified theories of gravity, which dispense with dark matter such as MOND and Verlinde's emergent gravity~\cite{XCOP1}.

%The cluster masses and errors were estimated for the X-COP sample using both the forward and backward method~\cite{Ettori2013}.
For all the clusters in the X-COP sample, the gas density has been obtained from the deprojected XMM surface brightness profile. The temperature is obtained from the deprojected XMM spectral estimates. The SZ (Sunyaev-Zel'dovich)~\cite{SZ} effect provides a direct measurement of the thermal electron pressure integrated along the line of sight. The pressure profile is recovered from the SZ signal measured in the all-sky survey by the \textit{Planck} SZ survey.~\cite{Planck_2015,xcop_thermoa2319}. The gas pressure has a smooth spatial distribution along the azimuth except for the clusters which are undergoing a merger. From  the ideal gas relation, one can deduce the pressure of the gas, when  only spectral measurements are available. Similarly the  temperature can be obtained using pressure obtained from deconvolution of \textit{Planck} measurements, when spectral measurements are not available~\cite{xcop_thermo}. Parametric temperature and pressure profiles for the XCOP samples can be found in ~\cite{xcop_thermo}.

Mass profiles have been calculated using multiple methods  for the X-COP sample, as documented in \citet{xcop_thermoa2319}.
The total mass and gas mass profiles, obtained using the \textit{backward} method (with NFW model as reference),  can be downloaded from the web, and are readily available in \texttt{.fits} files for all the clusters\footnote{X-COP Data: \url{https://dominiqueeckert.wixsite.com/xcop/data}}. The relative errors in mass measurements were less than 8\%~\cite{XCOP1}.
For our work, these were used to obtain the total mass.  To obtain the gas mass, the data in the files on the aforementioned website titled {\tt Gas mass and gas fraction profile}  were used. The gas density can be  obtained from $n_e$ using $\rho_{gas}={\mu} {m_p} (n_e+n_p)$, with $n_e=1.17 n_p$~\cite{XCOP1, xcop_fgas}. The electron number density ($n_e$), as mentioned earlier  has been estimated from the  deprojected XMM surface brightness profile. 

The stellar mass profiles are available for seven of the galaxy clusters~\cite{xcop_mstar}. Accordingly, we used them  for  these seven clusters, whereas  for the remaining five clusters, Eq.~\ref{eq:mstar} has been used to obtain the stellar mass. 
In order to obtain the (baryonic as well as total) mass estimates at any arbitrary radius, we  interpolate the different quantities using quadratic \textit{spline} interpolations, and then calculated their values at  100, 200, 400,  and 1000 kpc. Then, $a_{bar}$ is obtained using Eq.~\ref{eq:abar} and $a_{tot}$ is obtained using Eq.~\ref{eq:atotfrommass}. The  XCOP website also provides  $1\sigma$ errors for all the measurements, which we include for our analysis. We note that these errors do not account for non-thermal contributions to gas pressure and departures from spherical symmetry and hydrostatic equilibrium~\cite{XCOP1}.

To find the best-fit values for the linear regression relation between $\ln(a_{tot})$ and $\ln(a_{bar})$,  we
maximize the same likelihood as in Eq~\ref{eq:eq13}. Our best-fit values are shown in Table~\ref{tab:summary}. The 68\%, 90\%, and 99\%  confidence intervals for the slope, intercept, and natural log of intrinsic scatter are shown in Fig.~\ref{fig:xcop_rar_corner}. Both the slope and intercept differ from T20 by about $6.7\sigma$.   \rthis{We find an intrinsic scatter of about 21\%, which is of the same order of magnitude as found in T20. We also computed the residual scatter using the same method as in Sect.~\ref{sec:chandra}, by fitting a Gaussian distribution to difference in $\log_{10}$ of the total and baryonic accelerations and computing its standard deviation. For computation of the residual scatter for XCOP clusters, we binned the residuals into four bins. Then, residual scatter was calculated as the standard deviation of the best fit gaussian to the binned residuals.
We find that residual scatter computed using this method is equal to 0.11 dex, which is also comparable to T20. Therefore, we find that the X-COP sample obeys the RAR.}

\begin{figure}[H]
    \centering
    \includegraphics[width=0.5\textwidth]{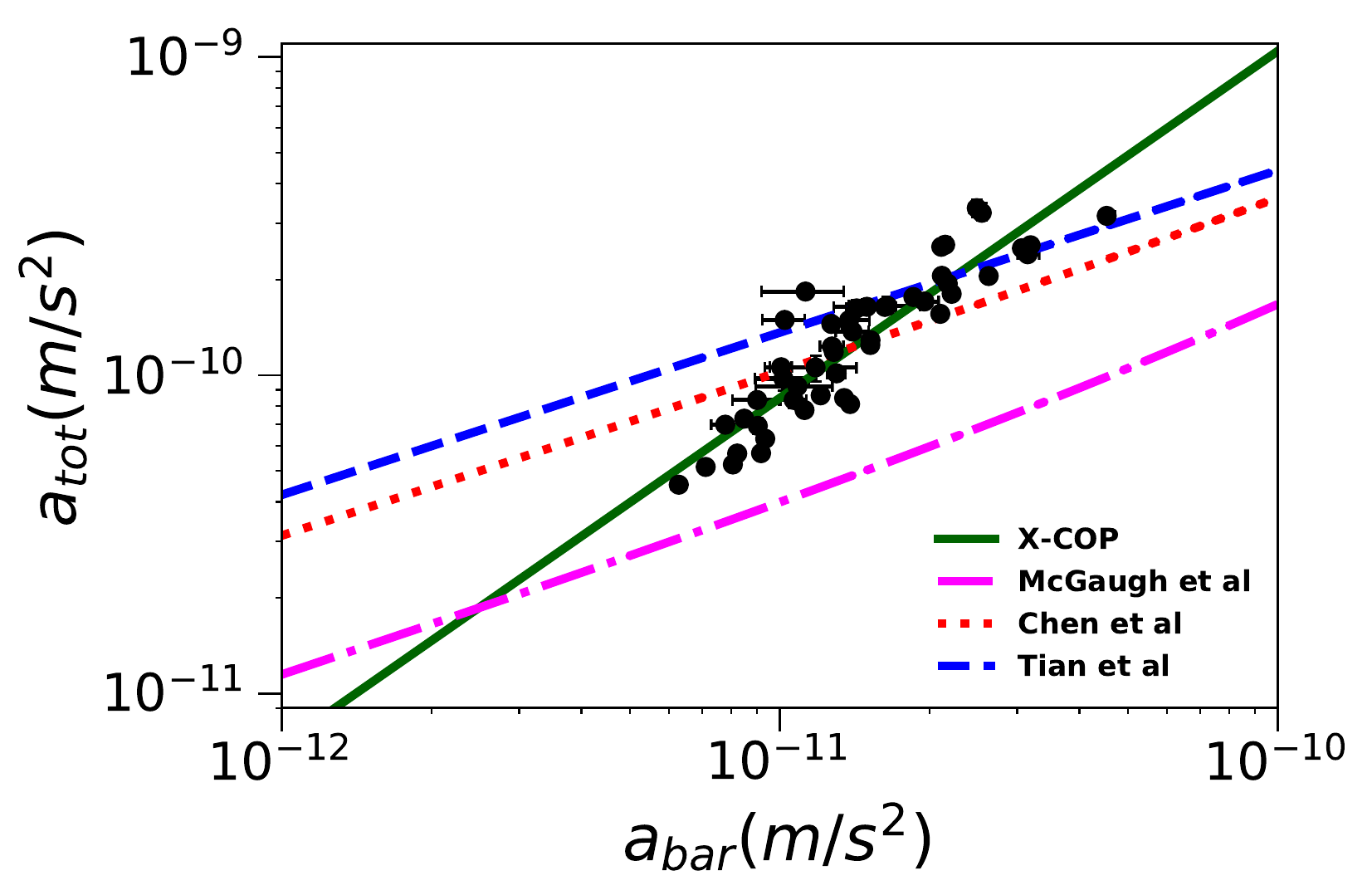}
    \caption{The best-fit straight line (green) for $a_{tot}$ vs $a_{bar}$ of the XCOP set of 12 clusters. The dots indicate the data points. The $a_{bar}$ and $a_{tot}$ have been plotted at 100, 200, 400, 1000 kpc for each of the clusters. The RAR fits from McGaugh et al~\cite{Mcgaugh16}, T20, CDP20 have been overlaid for reference. The RAR deduced from the  SPARC sample~\citep{Mcgaugh16} also does not agree with  the X-COP data.}
    \label{fig:xcop_rar}
\end{figure}

\begin{figure*}[h]
    \centering
    \includegraphics[width=15cm, height=13cm]{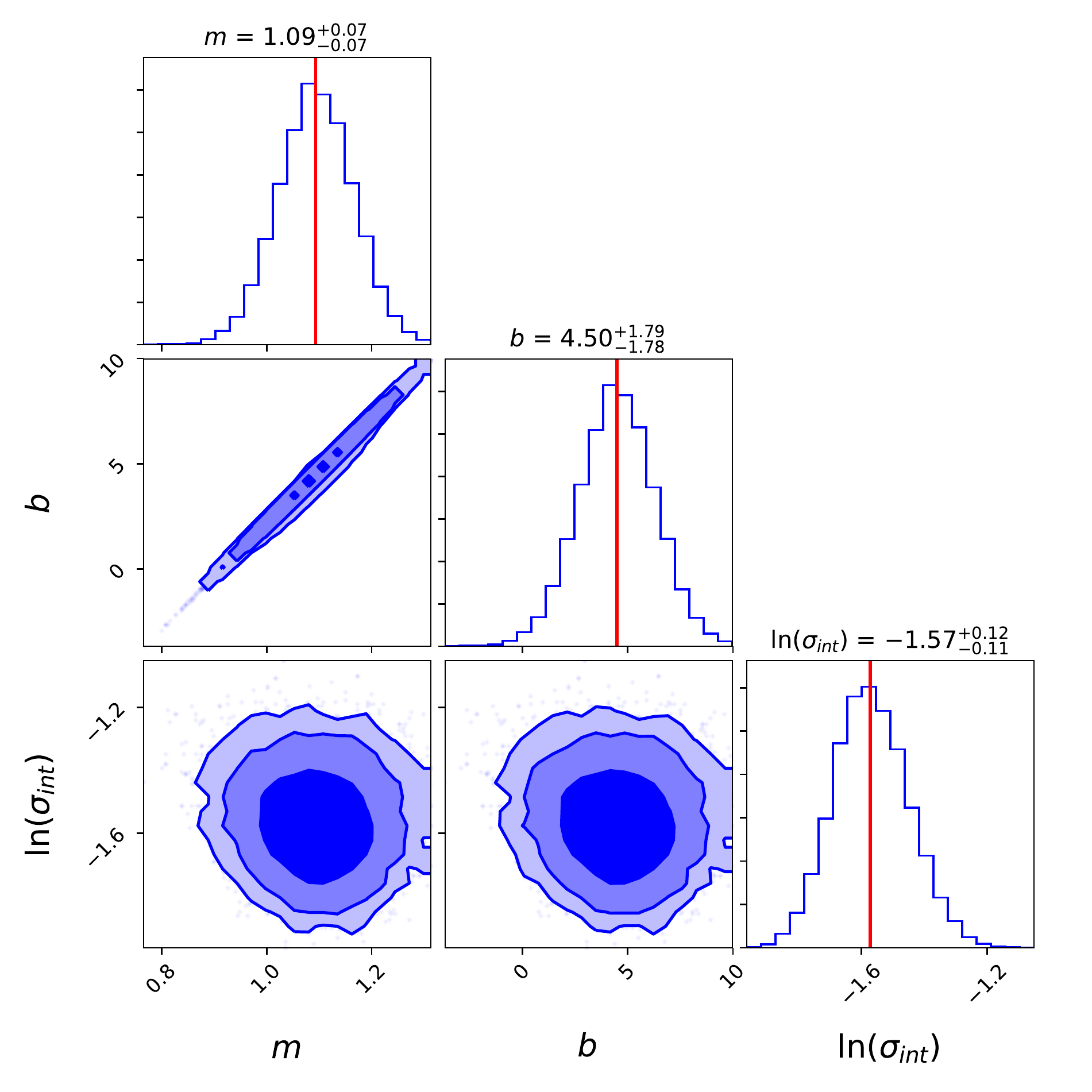}
    \caption{Plot showing 68\%, 90\% and 99\% marginalized credible intervals for the slope, intercept and  (ln) intrinsic scatter for the linear relation between $\ln a_{tot}$ and $\ln a_{bar}$ for the X-COP cluster sample.}
    \label{fig:xcop_rar_corner}
\end{figure*}

The best-fit RAR for our cluster sample along with those found in T20 and C20 are shown in Fig.~\ref{fig:xcop_rar}. We also show that the RAR found using the cluster sample.  If we fix the slope to 0.5, the best-fit value of $a_0$ is equal to $(1.12 \pm 0.11) \times 10^{-9} m/s^2$. This value is mid-way between the estimates in CDP20 and T20, and is also about an order of magnitude larger than that for galaxies.

%testing vs the plot I sent
%use pdf

\begin{table*}[t]
\begin{tabular}{|c|c|c|c|c|c|} \hline
Cluster Sample & Slope & Intercept & Intrinsic Scatter & Residual Scatter (dex) & $a_0 (m/s^2)$ \\ \hline
Chandra (this work) & $0.77\pm 0.1$ & $-3.5^{+2.6}_{-2.7}$ & ($0.0002 \pm 0.018$)\% & 0.14  &  $(9.26 \pm 1.66) \times 10^{-10}$   \\
X-COP (this work) & $1.09 _{- 0.07}^{+0.07}$& $4.50 _{- 1.78}^{ + 1.79}$ & $0.21_{ - 0.02}^{ + 0.03}$ & 0.11   &  $(1.12 \pm 0.11) \times 10^{-9}$    \\
CLASH~\cite{Tian} & $0.51^{0.04}_{-0.05}$ & $-9.8^{+1.07}_{-1.08}$ & $0.147^{+0.029}_{-0.028}$ & 0.11  &  $(2.02 \pm 0.11) \times 10^{-9}$    \\
HIFLUGCS  & $0.72_{- 0.06}^{ + 0.06}$ & $-5.49 _{- 1.39}^{ + 1.39 }$ & $0.27_{-0.02}^{+0.02}$  & 0.13  &  $9.5 \times 10^{-10}$  \\ \hline
\end{tabular}
\caption{\label{tab:summary}Summary of our results for linear regression between $\ln (a_{tot})$ as a function of $\ln (a_{bar})$ (Eq.~\ref{lineareq}),  for Chandra and  X-COP  samples analyzed,  as well as our re-analysis of the  HIFLUGCS sample. 
Also shown for comparison are the results from T20~\cite{Tian}.  \rthis{CDP20 obtain a higher residual scatter of 0.18 dex for this sample compared to our estimate of 0.13 dex.}
The value of  $a_0$ for the first three samples was obtained by assuming that the slope in the linear regression relation is equal to 0.5. For the HIFLUGCS sample, only $a_0$ shown here is from CDP20.}
\end{table*}

\section{Comparison of results}

In Sects.~\ref{sec:chandra} and \ref{sec:xcop}, we carried out a test of RAR in the same way as in T20 for the Chandra and X-COP cluster samples, respectively by fitting to Eq.~\ref{lineareq}. We have already compared our results for the slope and intercept  of these samples to those for the CLASH sample obtained in T20. However,  a direct comparison of our results with the analysis in CDP20  cannot be trivially done, since CDP20 used the original RAR relation  for spiral galaxies (Eq.~\ref{eq:RAR}), instead of Eq.~\ref{lineareq}.

Therefore, in order to bring the results from all the cluster samples to a common footing, we reanalyze the same dataset as in CDP20, using the same procedure as for the other samples.
We select a sample of 52 non-cool core clusters from the dataset analyzed in ~\cite{Chen}, by applying the condition that the cooling time, $t_{cool}>14$ Gyr~\citep{Chen} and $r_c \geq 100 r_c$, similar to CDP20. For the baryonic mass, we  also included the stellar mass using Eq. \ref{eq:mstar}. Then, we evaluated   $a_{tot}$ and $a_{bar}$  at 100, 200, 400, and 1000 kpc. The best-fit slope, intercept, and intrinsic scatter were then determined  using Eq.~\ref{lineareq}. Our best-fit values for the slope, intercept, and intrinsic scatter for this sample are shown in Table~\ref{tab:summary}. The residual scatter, which we get for this sample is 0.13 dex.
Note however that $a_0$ in this table for this sample is  the same as the estimate in CDP20. We get an  intrinsic scatter of about 27\%.   Our slope and intercept for the Chandra sample is consistent with the HIFLUGCS sample to within $1\sigma$. However, the X-COP slope and intercept are discrepant with respect to  our re-estimate of the HIFLUGCS sample  at  about 4$\sigma$ and 4.3$\sigma$ respectively.

\rthis{Therefore, we find that the X-COP sample and the CLASH sample analyzed in T20 show  comparable scatter, indicating that RAR is obeyed for the X-COP sample. However,  the HIFLUGS sample (analyzed in CDP20)  and the  Chandra samples  (analyzed in this work) either show a high intrinsic scatter in the RAR, or have large error bars for the slope and the intercept. However the slopes and intercepts for the X-COP and HIFLUGS sample are discrepant with respect to each other, which implies  that RAR is not universal for clusters.} 

\rthis{We now discuss, if the discrepancy between the CDP20 result  and the other analyses could be attributed to different assumptions.} CDP20 did not include the stellar mass in the baryonic mass budget. However, our re-analysis of CDP20, which includes the stellar mass contributions shows a residual scatter comparable to the CLASH and X-COP sample. Another difference between the T20 and the other three analyses is that T20 included the BCG stellar mass contribution. 
However, the contribution of BCG stellar mass  is dominant only at distances close to the cluster center,  whereas  the RAR analysis for the other three samples is done at distances far from the center, where the BCG stellar mass contribution is less than 1\%~\citep{Tian,Umetsu16}. 

\rthis{Nevertheless, T20 have found that the total stellar mass is important for the CLASH cluster sample within 200 kpc. If only the gas mass is used for their analysis, then the slope of the RAR is different from the estimated baryonic acceleration and the scatter will also be larger (Yong Tian, private communication). To investigate the effect of stellar mass, we redid our analysis for the X-COP sample (as it had smaller errors), without including the contribution of stellar mass to the total baryonic mass budget. We find that the new slope is given by $m=1.29 \pm 0.01$ and $b=9.33 \pm 0.01$, which is about $3\sigma$ (slope) and $2\sigma$  (intercept)  different, compared to  the inclusion of stellar mass. Without stellar mass, the intrinsic scatter which we obtain is 39\% and the residual scatter is 0.17 dex. Therefore, both the scatters are also increased if the stellar mass contribution is not included. We also computed the residual scatter in our re-analysis of HIFLUGS data, by excluding the stellar mass contribution. Upon doing that,  we get a residual scatter of 0.15 dex, based on the Gaussian standard deviation, which is larger than the 0.13 dex residual scatter which we obtain after inclusion of stellar mass. Therefore, we can conclude that excluding the stellar mass does increase the scatter and could be one of the reasons for the large scatter in CDP20.}

%Another  difference is that T20 used weak and strong lensing-based mass estimated to get the total mass. The remaining three analyses used X-ray hydrostatic masses. At this moment one cannot rule out that there are additional uncertainties related to hydrostatic masses (eg.~\cite{Biffi16}), which are causing the difference between T20 and the other results. A more definitive test can only be possible when a much  larger sample of X-ray masses in conjunction with lensing-based masses is used  to test the RAR. This should  be possible with the recent launch of the  eROSITA  X-ray satellite, which is expected to discover about 100,000 clusters.

\section{Conclusions}

Recently, two independent groups (CDP20 and T20) carried out a systematic test of a possible deterministic relation  between the baryonic and total acceleration  for galaxy clusters, known in the astrophysical literature as RAR. This was motivated by the very tight scatter found from a similar test done for spiral galaxies using the SPARC sample~\cite{Mcgaugh16}. T20 did this test with  the CLASH cluster sample using a combination of strong lensing, weak lensing and X-ray measurements, whereas CDP20 used  the hydrostatic masses of the non-cool core clusters from the ROSAT dataset. T20 found a tight empirical correlation between the two, with a slope consistent with the low-acceleration limit found for galaxies. However, they find an acceleration scale about ten times higher than for galaxies. However, CDP20 found  a large \rthis{residual  scatter about the RAR} of about 0.18 dex, and a higher acceleration scale about 4-10 times larger. Therefore, the CDP20 analysis shows that galaxy clusters do not obey the RAR.

Given the somewhat contradictory nature of the two results, we carry out an independent test of the same relation using two different cluster samples. The first sample is a dataset of 12 clusters imaged using Chandra telescope by Vikhlinin et al. The second sample which we use is the X-COP sample, in which a subset of 12 Planck-SZ clusters with a high SNR were imaged using the XMM-Newton telescope. For both these samples, we carry out a linear regression relation between the natural log of baryonic and the same for total acceleration. We also re-analyzed the HIFLUGCS cluster sample used in CDP20, using the same regression relation  used for the Chandra and X-COP sample, and also by including the stellar mass contribution in the baryonic mass budget. \rthis{We characterize the tightness of the RAR by computing both the intrinsic scatter and the residual scatter (in dex).}

For the Chandra sample, we find that the slope and intercept disagree with the estimates in T20 at about 2.2$\sigma$. However, they agree with the sample analyzed in CDP20 to within 1$\sigma$.  The residual scatter which we get is about 0.14 dex. The X-COP sample (which has  much smaller mass errors compared to the Chandra sample) disagrees with the estimates for both the slope and intercept, when compared to  the T20 and CDP20 estimates at about $6.7\sigma$ and $4\sigma$, respectively. \rthis{The residual scatter which we get for this sample is about the same as T20 (0.11 dex). The intrinsic scatter  which  we obtain for the X-COP sample  is about 20\%, which is about the same as that obtained in T20}

\rthis{Therefore,  we conclude that the X-COP sample  obeys the  RAR in agreement with the conclusions of  T20. The best-fit errors for the Chandra sample are somewhat large, and hence a decisive statement cannot be made for this sample.}

\rthis{However, given the differences in the  best fits for the slopes and intercepts between the different cluster samples, we find that  this relation is not universal for clusters.}

\rthis{We also find that exclusion of stellar mass changes the intercept and slope by about $2-3\sigma$, and also increases both the residual and intrinsic scatter. This may explain the high residual scatter seen in CDP20.}

If we assume that the total acceleration scales as the square root of the baryonic acceleration, then the characteristic acceleration scale we obtain is about the same order of magnitude as that obtained in T20 and CDP20, which is about an order of magnitude larger than what  found using the SPARC sample.

Looking to the future, a more definitive test should soon be possible with the recent launch of the eROSITA X-ray satellite~\cite{erosita}, which should provide X-ray  measurements for about 100,000 clusters, along with lensing and other mass estimates from follow-up programs,  which would allow high precision measurements of both the  baryonic as well as the total mass.

\begin{acknowledgements}
We are grateful to Stefano  Ettori and Dominik Eckert for providing the data and answering all our questions regarding the X-COP sample; Alexey Vikhlinin for the same regarding the Chandra sample;  Doug Edmonds, Yong Tian, Man-Ho Chan, and Antonio Del Popolo for helpful correspondence. The work of Sajal Gupta was supported by a DST-INSPIRE fellowship. We also thank the anonymous referee for several constructive comments on our manuscript.
\end{acknowledgements}
\bibliography{references}

\begin{thebibliography}{85}
\expandafter\ifx\csname natexlab\endcsname\relax\def\natexlab#1{#1}\fi
\expandafter\ifx\csname bibnamefont\endcsname\relax
  \def\bibnamefont#1{#1}\fi
\expandafter\ifx\csname bibfnamefont\endcsname\relax
  \def\bibfnamefont#1{#1}\fi
\expandafter\ifx\csname citenamefont\endcsname\relax
  \def\citenamefont#1{#1}\fi
\expandafter\ifx\csname url\endcsname\relax
  \def\url#1{\texttt{#1}}\fi
\expandafter\ifx\csname urlprefix\endcsname\relax\def\urlprefix{URL }\fi
\providecommand{\bibinfo}[2]{#2}
\providecommand{\eprint}[2][]{\url{#2}}

\bibitem[{\citenamefont{{Chan} and {Del Popolo}}(2020)}]{Chan20}
\bibinfo{author}{\bibfnamefont{M.~H.} \bibnamefont{{Chan}}} \bibnamefont{and}
  \bibinfo{author}{\bibfnamefont{A.}~\bibnamefont{{Del Popolo}}},
  \bibinfo{journal}{\mnras} p. \bibinfo{pages}{218} (\bibinfo{year}{2020}),
  \eprint{2001.06141}.

\bibitem[{\citenamefont{{Tian} et~al.}(2020{\natexlab{a}})\citenamefont{{Tian},
  {Umetsu}, {Ko}, {Donahue}, and {Chiu}}}]{Tian}
\bibinfo{author}{\bibfnamefont{Y.}~\bibnamefont{{Tian}}},
  \bibinfo{author}{\bibfnamefont{K.}~\bibnamefont{{Umetsu}}},
  \bibinfo{author}{\bibfnamefont{C.-M.} \bibnamefont{{Ko}}},
  \bibinfo{author}{\bibfnamefont{M.}~\bibnamefont{{Donahue}}},
  \bibnamefont{and} \bibinfo{author}{\bibfnamefont{I.~N.}
  \bibnamefont{{Chiu}}}, \bibinfo{journal}{\apj}
  \textbf{\bibinfo{volume}{896}}, \bibinfo{eid}{70}
  (\bibinfo{year}{2020}{\natexlab{a}}), \eprint{2001.08340}.

\bibitem[{\citenamefont{{McGaugh} et~al.}(2016)\citenamefont{{McGaugh},
  {Lelli}, and {Schombert}}}]{Mcgaugh16}
\bibinfo{author}{\bibfnamefont{S.~S.} \bibnamefont{{McGaugh}}},
  \bibinfo{author}{\bibfnamefont{F.}~\bibnamefont{{Lelli}}}, \bibnamefont{and}
  \bibinfo{author}{\bibfnamefont{J.~M.} \bibnamefont{{Schombert}}},
  \bibinfo{journal}{\prl} \textbf{\bibinfo{volume}{117}}, \bibinfo{eid}{201101}
  (\bibinfo{year}{2016}), \eprint{1609.05917}.

\bibitem[{\citenamefont{{Lelli} et~al.}(2016)\citenamefont{{Lelli}, {McGaugh},
  and {Schombert}}}]{Lelli}
\bibinfo{author}{\bibfnamefont{F.}~\bibnamefont{{Lelli}}},
  \bibinfo{author}{\bibfnamefont{S.~S.} \bibnamefont{{McGaugh}}},
  \bibnamefont{and} \bibinfo{author}{\bibfnamefont{J.~M.}
  \bibnamefont{{Schombert}}}, \bibinfo{journal}{\aj}
  \textbf{\bibinfo{volume}{152}}, \bibinfo{eid}{157} (\bibinfo{year}{2016}),
  \eprint{1606.09251}.

\bibitem[{\citenamefont{{Li} et~al.}(2018)\citenamefont{{Li}, {Lelli},
  {McGaugh}, and {Schombert}}}]{Li18}
\bibinfo{author}{\bibfnamefont{P.}~\bibnamefont{{Li}}},
  \bibinfo{author}{\bibfnamefont{F.}~\bibnamefont{{Lelli}}},
  \bibinfo{author}{\bibfnamefont{S.}~\bibnamefont{{McGaugh}}},
  \bibnamefont{and}
  \bibinfo{author}{\bibfnamefont{J.}~\bibnamefont{{Schombert}}},
  \bibinfo{journal}{\aap} \textbf{\bibinfo{volume}{615}}, \bibinfo{eid}{A3}
  (\bibinfo{year}{2018}), \eprint{1803.00022}.

\bibitem[{\citenamefont{{Chae} et~al.}(2020)\citenamefont{{Chae}, {Bernardi},
  {Dominguez Sanchez}, and {Sheth}}}]{Sheth}
\bibinfo{author}{\bibfnamefont{K.-H.} \bibnamefont{{Chae}}},
  \bibinfo{author}{\bibfnamefont{M.}~\bibnamefont{{Bernardi}}},
  \bibinfo{author}{\bibfnamefont{H.}~\bibnamefont{{Dominguez Sanchez}}},
  \bibnamefont{and} \bibinfo{author}{\bibfnamefont{R.~K.}
  \bibnamefont{{Sheth}}}, \bibinfo{journal}{arXiv e-prints}
  \bibinfo{eid}{arXiv:2010.10779} (\bibinfo{year}{2020}), \eprint{2010.10779}.

\bibitem[{\citenamefont{{McGaugh}}(2004)}]{Mcgaugh04}
\bibinfo{author}{\bibfnamefont{S.~S.} \bibnamefont{{McGaugh}}},
  \bibinfo{journal}{\apj} \textbf{\bibinfo{volume}{609}}, \bibinfo{pages}{652}
  (\bibinfo{year}{2004}), \eprint{astro-ph/0403610}.

\bibitem[{\citenamefont{{Famaey} and {McGaugh}}(2012)}]{Famaey12}
\bibinfo{author}{\bibfnamefont{B.}~\bibnamefont{{Famaey}}} \bibnamefont{and}
  \bibinfo{author}{\bibfnamefont{S.~S.} \bibnamefont{{McGaugh}}},
  \bibinfo{journal}{Living Reviews in Relativity}
  \textbf{\bibinfo{volume}{15}}, \bibinfo{eid}{10} (\bibinfo{year}{2012}),
  \eprint{1112.3960}.

\bibitem[{\citenamefont{{Milgrom}}(2016)}]{Milgrom16}
\bibinfo{author}{\bibfnamefont{M.}~\bibnamefont{{Milgrom}}},
  \bibinfo{journal}{arXiv e-prints} \bibinfo{eid}{arXiv:1609.06642}
  (\bibinfo{year}{2016}), \eprint{1609.06642}.

\bibitem[{\citenamefont{{Lelli}
  et~al.}(2017{\natexlab{a}})\citenamefont{{Lelli}, {McGaugh}, {Schombert}, and
  {Pawlowski}}}]{Lelli17}
\bibinfo{author}{\bibfnamefont{F.}~\bibnamefont{{Lelli}}},
  \bibinfo{author}{\bibfnamefont{S.~S.} \bibnamefont{{McGaugh}}},
  \bibinfo{author}{\bibfnamefont{J.~M.} \bibnamefont{{Schombert}}},
  \bibnamefont{and} \bibinfo{author}{\bibfnamefont{M.~S.}
  \bibnamefont{{Pawlowski}}}, \bibinfo{journal}{\apj}
  \textbf{\bibinfo{volume}{836}}, \bibinfo{eid}{152}
  (\bibinfo{year}{2017}{\natexlab{a}}), \eprint{1610.08981}.

\bibitem[{\citenamefont{{Wheeler} et~al.}(2019)\citenamefont{{Wheeler},
  {Hopkins}, and {Dor{\'e}}}}]{Wheeler}
\bibinfo{author}{\bibfnamefont{C.}~\bibnamefont{{Wheeler}}},
  \bibinfo{author}{\bibfnamefont{P.~F.} \bibnamefont{{Hopkins}}},
  \bibnamefont{and}
  \bibinfo{author}{\bibfnamefont{O.}~\bibnamefont{{Dor{\'e}}}},
  \bibinfo{journal}{\apj} \textbf{\bibinfo{volume}{882}}, \bibinfo{eid}{46}
  (\bibinfo{year}{2019}), \eprint{1803.01849}.

\bibitem[{\citenamefont{{Rodrigues}
  et~al.}(2018{\natexlab{a}})\citenamefont{{Rodrigues}, {Marra}, {del Popolo},
  and {Davari}}}]{Rodrigues}
\bibinfo{author}{\bibfnamefont{D.~C.} \bibnamefont{{Rodrigues}}},
  \bibinfo{author}{\bibfnamefont{V.}~\bibnamefont{{Marra}}},
  \bibinfo{author}{\bibfnamefont{A.}~\bibnamefont{{del Popolo}}},
  \bibnamefont{and} \bibinfo{author}{\bibfnamefont{Z.}~\bibnamefont{{Davari}}},
  \bibinfo{journal}{Nature Astronomy} \textbf{\bibinfo{volume}{2}},
  \bibinfo{pages}{668} (\bibinfo{year}{2018}{\natexlab{a}}),
  \eprint{1806.06803}.

\bibitem[{\citenamefont{{Zobnina} and {Zasov}}(2020)}]{Zasov}
\bibinfo{author}{\bibfnamefont{D.~I.} \bibnamefont{{Zobnina}}}
  \bibnamefont{and} \bibinfo{author}{\bibfnamefont{A.~V.}
  \bibnamefont{{Zasov}}}, \bibinfo{journal}{Astronomy Reports}
  \textbf{\bibinfo{volume}{64}}, \bibinfo{pages}{295} (\bibinfo{year}{2020}),
  \eprint{2003.08845}.

\bibitem[{\citenamefont{{Marra} et~al.}(2020)\citenamefont{{Marra},
  {Rodrigues}, and {de Almeida}}}]{Marra}
\bibinfo{author}{\bibfnamefont{V.}~\bibnamefont{{Marra}}},
  \bibinfo{author}{\bibfnamefont{D.~C.} \bibnamefont{{Rodrigues}}},
  \bibnamefont{and} \bibinfo{author}{\bibfnamefont{{\'A}.~O.~F.}
  \bibnamefont{{de Almeida}}}, \bibinfo{journal}{\mnras}
  \textbf{\bibinfo{volume}{494}}, \bibinfo{pages}{2875} (\bibinfo{year}{2020}),
  \eprint{2002.03946}.

\bibitem[{\citenamefont{{Zhou} et~al.}(2020)\citenamefont{{Zhou}, {Del Popolo},
  and {Chang}}}]{Zhu20}
\bibinfo{author}{\bibfnamefont{Y.}~\bibnamefont{{Zhou}}},
  \bibinfo{author}{\bibfnamefont{A.}~\bibnamefont{{Del Popolo}}},
  \bibnamefont{and} \bibinfo{author}{\bibfnamefont{Z.}~\bibnamefont{{Chang}}},
  \bibinfo{journal}{Physics of the Dark Universe}
  \textbf{\bibinfo{volume}{28}}, \bibinfo{eid}{100468} (\bibinfo{year}{2020}),
  \eprint{2008.04065}.

\bibitem[{\citenamefont{{McGaugh} et~al.}(2018)\citenamefont{{McGaugh}, {Li},
  {Lelli}, and {Schombert}}}]{McGaughNature}
\bibinfo{author}{\bibfnamefont{S.~S.} \bibnamefont{{McGaugh}}},
  \bibinfo{author}{\bibfnamefont{P.}~\bibnamefont{{Li}}},
  \bibinfo{author}{\bibfnamefont{F.}~\bibnamefont{{Lelli}}}, \bibnamefont{and}
  \bibinfo{author}{\bibfnamefont{J.~M.} \bibnamefont{{Schombert}}},
  \bibinfo{journal}{Nature Astronomy} \textbf{\bibinfo{volume}{2}},
  \bibinfo{pages}{924} (\bibinfo{year}{2018}).

\bibitem[{\citenamefont{{Kroupa} et~al.}(2018)\citenamefont{{Kroupa}, {Banik},
  {Haghi}, {Zonoozi}, {Dabringhausen}, {Javanmardi}, {M{\"u}ller}, {Wu}, and
  {Zhao}}}]{Kroupa}
\bibinfo{author}{\bibfnamefont{P.}~\bibnamefont{{Kroupa}}},
  \bibinfo{author}{\bibfnamefont{I.}~\bibnamefont{{Banik}}},
  \bibinfo{author}{\bibfnamefont{H.}~\bibnamefont{{Haghi}}},
  \bibinfo{author}{\bibfnamefont{A.~H.} \bibnamefont{{Zonoozi}}},
  \bibinfo{author}{\bibfnamefont{J.}~\bibnamefont{{Dabringhausen}}},
  \bibinfo{author}{\bibfnamefont{B.}~\bibnamefont{{Javanmardi}}},
  \bibinfo{author}{\bibfnamefont{O.}~\bibnamefont{{M{\"u}ller}}},
  \bibinfo{author}{\bibfnamefont{X.}~\bibnamefont{{Wu}}}, \bibnamefont{and}
  \bibinfo{author}{\bibfnamefont{H.}~\bibnamefont{{Zhao}}},
  \bibinfo{journal}{Nature Astronomy} \textbf{\bibinfo{volume}{2}},
  \bibinfo{pages}{925} (\bibinfo{year}{2018}), \eprint{1811.11754}.

\bibitem[{\citenamefont{{Rodrigues}
  et~al.}(2018{\natexlab{b}})\citenamefont{{Rodrigues}, {Marra}, {Del Popolo},
  and {Davari}}}]{delpopolo18}
\bibinfo{author}{\bibfnamefont{D.~C.} \bibnamefont{{Rodrigues}}},
  \bibinfo{author}{\bibfnamefont{V.}~\bibnamefont{{Marra}}},
  \bibinfo{author}{\bibfnamefont{A.}~\bibnamefont{{Del Popolo}}},
  \bibnamefont{and} \bibinfo{author}{\bibfnamefont{Z.}~\bibnamefont{{Davari}}},
  \bibinfo{journal}{Nature Astronomy} \textbf{\bibinfo{volume}{2}},
  \bibinfo{pages}{927} (\bibinfo{year}{2018}{\natexlab{b}}),
  \eprint{1811.05882}.

\bibitem[{\citenamefont{{Rodrigues} et~al.}(2020)\citenamefont{{Rodrigues},
  {Marra}, {Del Popolo}, and {Davari}}}]{Rodrigues20}
\bibinfo{author}{\bibfnamefont{D.~C.} \bibnamefont{{Rodrigues}}},
  \bibinfo{author}{\bibfnamefont{V.}~\bibnamefont{{Marra}}},
  \bibinfo{author}{\bibfnamefont{A.}~\bibnamefont{{Del Popolo}}},
  \bibnamefont{and} \bibinfo{author}{\bibfnamefont{Z.}~\bibnamefont{{Davari}}},
  \bibinfo{journal}{Nature Astronomy} \textbf{\bibinfo{volume}{4}},
  \bibinfo{pages}{134} (\bibinfo{year}{2020}), \eprint{2002.01970}.

\bibitem[{\citenamefont{{Desmond}}(2017)}]{Desmond}
\bibinfo{author}{\bibfnamefont{H.}~\bibnamefont{{Desmond}}},
  \bibinfo{journal}{\mnras} \textbf{\bibinfo{volume}{464}},
  \bibinfo{pages}{4160} (\bibinfo{year}{2017}), \eprint{1607.01800}.

\bibitem[{\citenamefont{{Navarro} et~al.}(2017)\citenamefont{{Navarro},
  {Ben{\'\i}tez-Llambay}, {Fattahi}, {Frenk}, {Ludlow}, {Oman}, {Schaller}, and
  {Theuns}}}]{Navarro}
\bibinfo{author}{\bibfnamefont{J.~F.} \bibnamefont{{Navarro}}},
  \bibinfo{author}{\bibfnamefont{A.}~\bibnamefont{{Ben{\'\i}tez-Llambay}}},
  \bibinfo{author}{\bibfnamefont{A.}~\bibnamefont{{Fattahi}}},
  \bibinfo{author}{\bibfnamefont{C.~S.} \bibnamefont{{Frenk}}},
  \bibinfo{author}{\bibfnamefont{A.~D.} \bibnamefont{{Ludlow}}},
  \bibinfo{author}{\bibfnamefont{K.~A.} \bibnamefont{{Oman}}},
  \bibinfo{author}{\bibfnamefont{M.}~\bibnamefont{{Schaller}}},
  \bibnamefont{and} \bibinfo{author}{\bibfnamefont{T.}~\bibnamefont{{Theuns}}},
  \bibinfo{journal}{\mnras} \textbf{\bibinfo{volume}{471}},
  \bibinfo{pages}{1841} (\bibinfo{year}{2017}), \eprint{1612.06329}.

\bibitem[{\citenamefont{{Dai} and {Lu}}(2017)}]{Lu}
\bibinfo{author}{\bibfnamefont{D.-C.} \bibnamefont{{Dai}}} \bibnamefont{and}
  \bibinfo{author}{\bibfnamefont{C.}~\bibnamefont{{Lu}}},
  \bibinfo{journal}{\prd} \textbf{\bibinfo{volume}{96}}, \bibinfo{eid}{124016}
  (\bibinfo{year}{2017}), \eprint{1712.01654}.

\bibitem[{\citenamefont{{Keller} and {Wadsley}}(2017)}]{Keller}
\bibinfo{author}{\bibfnamefont{B.~W.} \bibnamefont{{Keller}}} \bibnamefont{and}
  \bibinfo{author}{\bibfnamefont{J.~W.} \bibnamefont{{Wadsley}}},
  \bibinfo{journal}{\apjl} \textbf{\bibinfo{volume}{835}}, \bibinfo{eid}{L17}
  (\bibinfo{year}{2017}), \eprint{1610.06183}.

\bibitem[{\citenamefont{{Garaldi} et~al.}(2018)\citenamefont{{Garaldi},
  {Romano-D{\'\i}az}, {Porciani}, and {Pawlowski}}}]{Garaldi}
\bibinfo{author}{\bibfnamefont{E.}~\bibnamefont{{Garaldi}}},
  \bibinfo{author}{\bibfnamefont{E.}~\bibnamefont{{Romano-D{\'\i}az}}},
  \bibinfo{author}{\bibfnamefont{C.}~\bibnamefont{{Porciani}}},
  \bibnamefont{and} \bibinfo{author}{\bibfnamefont{M.~S.}
  \bibnamefont{{Pawlowski}}}, \bibinfo{journal}{\prl}
  \textbf{\bibinfo{volume}{120}}, \bibinfo{eid}{261301} (\bibinfo{year}{2018}),
  \eprint{1712.04448}.

\bibitem[{\citenamefont{{Ludlow} et~al.}(2017)\citenamefont{{Ludlow},
  {Ben{\'\i}tez-Llambay}, {Schaller}, {Theuns}, {Frenk}, {Bower}, {Schaye},
  {Crain}, {Navarro}, {Fattahi} et~al.}}]{Ludlow}
\bibinfo{author}{\bibfnamefont{A.~D.} \bibnamefont{{Ludlow}}},
  \bibinfo{author}{\bibfnamefont{A.}~\bibnamefont{{Ben{\'\i}tez-Llambay}}},
  \bibinfo{author}{\bibfnamefont{M.}~\bibnamefont{{Schaller}}},
  \bibinfo{author}{\bibfnamefont{T.}~\bibnamefont{{Theuns}}},
  \bibinfo{author}{\bibfnamefont{C.~S.} \bibnamefont{{Frenk}}},
  \bibinfo{author}{\bibfnamefont{R.}~\bibnamefont{{Bower}}},
  \bibinfo{author}{\bibfnamefont{J.}~\bibnamefont{{Schaye}}},
  \bibinfo{author}{\bibfnamefont{R.~A.} \bibnamefont{{Crain}}},
  \bibinfo{author}{\bibfnamefont{J.~F.} \bibnamefont{{Navarro}}},
  \bibinfo{author}{\bibfnamefont{A.}~\bibnamefont{{Fattahi}}},
  \bibnamefont{et~al.}, \bibinfo{journal}{\prl} \textbf{\bibinfo{volume}{118}},
  \bibinfo{eid}{161103} (\bibinfo{year}{2017}), \eprint{1610.07663}.

\bibitem[{\citenamefont{{Stone} and {Courteau}}(2019)}]{Stone}
\bibinfo{author}{\bibfnamefont{C.}~\bibnamefont{{Stone}}} \bibnamefont{and}
  \bibinfo{author}{\bibfnamefont{S.}~\bibnamefont{{Courteau}}},
  \bibinfo{journal}{\apj} \textbf{\bibinfo{volume}{882}}, \bibinfo{eid}{6}
  (\bibinfo{year}{2019}), \eprint{1908.06105}.

\bibitem[{\citenamefont{{Ren} et~al.}(2019)\citenamefont{{Ren}, {Kwa},
  {Kaplinghat}, and {Yu}}}]{Ren}
\bibinfo{author}{\bibfnamefont{T.}~\bibnamefont{{Ren}}},
  \bibinfo{author}{\bibfnamefont{A.}~\bibnamefont{{Kwa}}},
  \bibinfo{author}{\bibfnamefont{M.}~\bibnamefont{{Kaplinghat}}},
  \bibnamefont{and} \bibinfo{author}{\bibfnamefont{H.-B.} \bibnamefont{{Yu}}},
  \bibinfo{journal}{Physical Review X} \textbf{\bibinfo{volume}{9}},
  \bibinfo{eid}{031020} (\bibinfo{year}{2019}), \eprint{1808.05695}.

\bibitem[{\citenamefont{{Berezhiani} et~al.}(2018)\citenamefont{{Berezhiani},
  {Famaey}, and {Khoury}}}]{Khoury}
\bibinfo{author}{\bibfnamefont{L.}~\bibnamefont{{Berezhiani}}},
  \bibinfo{author}{\bibfnamefont{B.}~\bibnamefont{{Famaey}}}, \bibnamefont{and}
  \bibinfo{author}{\bibfnamefont{J.}~\bibnamefont{{Khoury}}},
  \bibinfo{journal}{\jcap} \textbf{\bibinfo{volume}{2018}}, \bibinfo{eid}{021}
  (\bibinfo{year}{2018}), \eprint{1711.05748}.

\bibitem[{\citenamefont{{Green} and {Moffat}}(2019)}]{Green}
\bibinfo{author}{\bibfnamefont{M.~A.} \bibnamefont{{Green}}} \bibnamefont{and}
  \bibinfo{author}{\bibfnamefont{J.~W.} \bibnamefont{{Moffat}}},
  \bibinfo{journal}{Physics of the Dark Universe}
  \textbf{\bibinfo{volume}{25}}, \bibinfo{eid}{100323} (\bibinfo{year}{2019}),
  \eprint{1905.09476}.

\bibitem[{\citenamefont{{O'Brien} et~al.}(2018)\citenamefont{{O'Brien},
  {Chiarelli}, and {Mannheim}}}]{Mannheim}
\bibinfo{author}{\bibfnamefont{J.~G.} \bibnamefont{{O'Brien}}},
  \bibinfo{author}{\bibfnamefont{T.~L.} \bibnamefont{{Chiarelli}}},
  \bibnamefont{and} \bibinfo{author}{\bibfnamefont{P.~D.}
  \bibnamefont{{Mannheim}}}, \bibinfo{journal}{Physics Letters B}
  \textbf{\bibinfo{volume}{782}}, \bibinfo{pages}{433} (\bibinfo{year}{2018}),
  \eprint{1704.03921}.

\bibitem[{\citenamefont{{Lelli}
  et~al.}(2017{\natexlab{b}})\citenamefont{{Lelli}, {McGaugh}, and
  {Schombert}}}]{Schombert}
\bibinfo{author}{\bibfnamefont{F.}~\bibnamefont{{Lelli}}},
  \bibinfo{author}{\bibfnamefont{S.~S.} \bibnamefont{{McGaugh}}},
  \bibnamefont{and} \bibinfo{author}{\bibfnamefont{J.~M.}
  \bibnamefont{{Schombert}}}, \bibinfo{journal}{\mnras}
  \textbf{\bibinfo{volume}{468}}, \bibinfo{pages}{L68}
  (\bibinfo{year}{2017}{\natexlab{b}}), \eprint{1702.04355}.

\bibitem[{\citenamefont{{Voit}}(2005)}]{Voit}
\bibinfo{author}{\bibfnamefont{G.~M.} \bibnamefont{{Voit}}},
  \bibinfo{journal}{Reviews of Modern Physics} \textbf{\bibinfo{volume}{77}},
  \bibinfo{pages}{207} (\bibinfo{year}{2005}), \eprint{astro-ph/0410173}.

\bibitem[{\citenamefont{{Vikhlinin} et~al.}(2014)\citenamefont{{Vikhlinin},
  {Kravtsov}, {Markevich}, {Sunyaev}, and {Churazov}}}]{Vikhlininrev}
\bibinfo{author}{\bibfnamefont{A.~A.} \bibnamefont{{Vikhlinin}}},
  \bibinfo{author}{\bibfnamefont{A.~V.} \bibnamefont{{Kravtsov}}},
  \bibinfo{author}{\bibfnamefont{M.~L.} \bibnamefont{{Markevich}}},
  \bibinfo{author}{\bibfnamefont{R.~A.} \bibnamefont{{Sunyaev}}},
  \bibnamefont{and} \bibinfo{author}{\bibfnamefont{E.~M.}
  \bibnamefont{{Churazov}}}, \bibinfo{journal}{Physics Uspekhi}
  \textbf{\bibinfo{volume}{57}}, \bibinfo{eid}{317-341} (\bibinfo{year}{2014}).

\bibitem[{\citenamefont{{Allen} et~al.}(2011)\citenamefont{{Allen}, {Evrard},
  and {Mantz}}}]{Allen}
\bibinfo{author}{\bibfnamefont{S.~W.} \bibnamefont{{Allen}}},
  \bibinfo{author}{\bibfnamefont{A.~E.} \bibnamefont{{Evrard}}},
  \bibnamefont{and} \bibinfo{author}{\bibfnamefont{A.~B.}
  \bibnamefont{{Mantz}}}, \bibinfo{journal}{\araa}
  \textbf{\bibinfo{volume}{49}}, \bibinfo{pages}{409} (\bibinfo{year}{2011}),
  \eprint{1103.4829}.

\bibitem[{\citenamefont{{Desai}}(2018)}]{Desai18}
\bibinfo{author}{\bibfnamefont{S.}~\bibnamefont{{Desai}}},
  \bibinfo{journal}{Physics Letters B} \textbf{\bibinfo{volume}{778}},
  \bibinfo{pages}{325} (\bibinfo{year}{2018}), \eprint{1708.06502}.

\bibitem[{\citenamefont{{The} and {White}}(1988)}]{White88}
\bibinfo{author}{\bibfnamefont{L.~S.} \bibnamefont{{The}}} \bibnamefont{and}
  \bibinfo{author}{\bibfnamefont{S.~D.~M.} \bibnamefont{{White}}},
  \bibinfo{journal}{\aj} \textbf{\bibinfo{volume}{95}}, \bibinfo{pages}{1642}
  (\bibinfo{year}{1988}).

\bibitem[{\citenamefont{{Gerbal} et~al.}(1992)\citenamefont{{Gerbal}, {Durret},
  {Lachieze-Rey}, and {Lima-Neto}}}]{Gerbal}
\bibinfo{author}{\bibfnamefont{D.}~\bibnamefont{{Gerbal}}},
  \bibinfo{author}{\bibfnamefont{F.}~\bibnamefont{{Durret}}},
  \bibinfo{author}{\bibfnamefont{M.}~\bibnamefont{{Lachieze-Rey}}},
  \bibnamefont{and}
  \bibinfo{author}{\bibfnamefont{G.}~\bibnamefont{{Lima-Neto}}},
  \bibinfo{journal}{\aap} \textbf{\bibinfo{volume}{262}}, \bibinfo{pages}{395}
  (\bibinfo{year}{1992}).

\bibitem[{\citenamefont{{Aguirre} et~al.}(2001)\citenamefont{{Aguirre},
  {Schaye}, and {Quataert}}}]{Aguirre}
\bibinfo{author}{\bibfnamefont{A.}~\bibnamefont{{Aguirre}}},
  \bibinfo{author}{\bibfnamefont{J.}~\bibnamefont{{Schaye}}}, \bibnamefont{and}
  \bibinfo{author}{\bibfnamefont{E.}~\bibnamefont{{Quataert}}},
  \bibinfo{journal}{\apj} \textbf{\bibinfo{volume}{561}}, \bibinfo{pages}{550}
  (\bibinfo{year}{2001}), \eprint{astro-ph/0105184}.

\bibitem[{\citenamefont{{Sanders}}(2003)}]{Sanders}
\bibinfo{author}{\bibfnamefont{R.~H.} \bibnamefont{{Sanders}}},
  \bibinfo{journal}{\mnras} \textbf{\bibinfo{volume}{342}},
  \bibinfo{pages}{901} (\bibinfo{year}{2003}), \eprint{astro-ph/0212293}.

\bibitem[{\citenamefont{{Pointecouteau} and {Silk}}(2005)}]{Silk}
\bibinfo{author}{\bibfnamefont{E.}~\bibnamefont{{Pointecouteau}}}
  \bibnamefont{and} \bibinfo{author}{\bibfnamefont{J.}~\bibnamefont{{Silk}}},
  \bibinfo{journal}{\mnras} \textbf{\bibinfo{volume}{364}},
  \bibinfo{pages}{654} (\bibinfo{year}{2005}), \eprint{astro-ph/0505017}.

\bibitem[{\citenamefont{{Angus} et~al.}(2008)\citenamefont{{Angus}, {Famaey},
  and {Buote}}}]{Angus}
\bibinfo{author}{\bibfnamefont{G.~W.} \bibnamefont{{Angus}}},
  \bibinfo{author}{\bibfnamefont{B.}~\bibnamefont{{Famaey}}}, \bibnamefont{and}
  \bibinfo{author}{\bibfnamefont{D.~A.} \bibnamefont{{Buote}}},
  \bibinfo{journal}{\mnras} \textbf{\bibinfo{volume}{387}},
  \bibinfo{pages}{1470} (\bibinfo{year}{2008}), \eprint{0709.0108}.

\bibitem[{\citenamefont{{Natarajan} and {Zhao}}(2008)}]{Natarajan}
\bibinfo{author}{\bibfnamefont{P.}~\bibnamefont{{Natarajan}}} \bibnamefont{and}
  \bibinfo{author}{\bibfnamefont{H.}~\bibnamefont{{Zhao}}},
  \bibinfo{journal}{\mnras} \textbf{\bibinfo{volume}{389}},
  \bibinfo{pages}{250} (\bibinfo{year}{2008}), \eprint{0806.3080}.

\bibitem[{\citenamefont{{Dodelson}}(2011)}]{Dodelson}
\bibinfo{author}{\bibfnamefont{S.}~\bibnamefont{{Dodelson}}},
  \bibinfo{journal}{International Journal of Modern Physics D}
  \textbf{\bibinfo{volume}{20}}, \bibinfo{pages}{2749} (\bibinfo{year}{2011}),
  \eprint{1112.1320}.

\bibitem[{\citenamefont{{Boran} et~al.}(2018)\citenamefont{{Boran}, {Desai},
  {Kahya}, and {Woodard}}}]{Woodard}
\bibinfo{author}{\bibfnamefont{S.}~\bibnamefont{{Boran}}},
  \bibinfo{author}{\bibfnamefont{S.}~\bibnamefont{{Desai}}},
  \bibinfo{author}{\bibfnamefont{E.~O.} \bibnamefont{{Kahya}}},
  \bibnamefont{and} \bibinfo{author}{\bibfnamefont{R.~P.}
  \bibnamefont{{Woodard}}}, \bibinfo{journal}{\prd}
  \textbf{\bibinfo{volume}{97}}, \bibinfo{eid}{041501} (\bibinfo{year}{2018}),
  \eprint{1710.06168}.

\bibitem[{\citenamefont{{Famaey} et~al.}(2018)\citenamefont{{Famaey}, {Khoury},
  and {Penco}}}]{Famaey18}
\bibinfo{author}{\bibfnamefont{B.}~\bibnamefont{{Famaey}}},
  \bibinfo{author}{\bibfnamefont{J.}~\bibnamefont{{Khoury}}}, \bibnamefont{and}
  \bibinfo{author}{\bibfnamefont{R.}~\bibnamefont{{Penco}}},
  \bibinfo{journal}{\jcap} \textbf{\bibinfo{volume}{2018}}, \bibinfo{eid}{038}
  (\bibinfo{year}{2018}), \eprint{1712.01316}.

\bibitem[{\citenamefont{{Chen} et~al.}(2007)\citenamefont{{Chen}, {Reiprich},
  {B{\"o}hringer}, {Ikebe}, and {Zhang}}}]{Chen}
\bibinfo{author}{\bibfnamefont{Y.}~\bibnamefont{{Chen}}},
  \bibinfo{author}{\bibfnamefont{T.~H.} \bibnamefont{{Reiprich}}},
  \bibinfo{author}{\bibfnamefont{H.}~\bibnamefont{{B{\"o}hringer}}},
  \bibinfo{author}{\bibfnamefont{Y.}~\bibnamefont{{Ikebe}}}, \bibnamefont{and}
  \bibinfo{author}{\bibfnamefont{Y.~Y.} \bibnamefont{{Zhang}}},
  \bibinfo{journal}{\aap} \textbf{\bibinfo{volume}{466}}, \bibinfo{pages}{805}
  (\bibinfo{year}{2007}), \eprint{astro-ph/0702482}.

\bibitem[{\citenamefont{{Cavaliere} and {Fusco-Femiano}}(1978)}]{betamodel}
\bibinfo{author}{\bibfnamefont{A.}~\bibnamefont{{Cavaliere}}} \bibnamefont{and}
  \bibinfo{author}{\bibfnamefont{R.}~\bibnamefont{{Fusco-Femiano}}},
  \bibinfo{journal}{\aap} \textbf{\bibinfo{volume}{70}}, \bibinfo{pages}{677}
  (\bibinfo{year}{1978}).

\bibitem[{\citenamefont{{Postman} et~al.}(2012)\citenamefont{{Postman}, {Coe},
  {Ben{\'\i}tez}, {Bradley}, {Broadhurst}, {Donahue}, {Ford}, {Graur},
  {Graves}, {Jouvel} et~al.}}]{Postman}
\bibinfo{author}{\bibfnamefont{M.}~\bibnamefont{{Postman}}},
  \bibinfo{author}{\bibfnamefont{D.}~\bibnamefont{{Coe}}},
  \bibinfo{author}{\bibfnamefont{N.}~\bibnamefont{{Ben{\'\i}tez}}},
  \bibinfo{author}{\bibfnamefont{L.}~\bibnamefont{{Bradley}}},
  \bibinfo{author}{\bibfnamefont{T.}~\bibnamefont{{Broadhurst}}},
  \bibinfo{author}{\bibfnamefont{M.}~\bibnamefont{{Donahue}}},
  \bibinfo{author}{\bibfnamefont{H.}~\bibnamefont{{Ford}}},
  \bibinfo{author}{\bibfnamefont{O.}~\bibnamefont{{Graur}}},
  \bibinfo{author}{\bibfnamefont{G.}~\bibnamefont{{Graves}}},
  \bibinfo{author}{\bibfnamefont{S.}~\bibnamefont{{Jouvel}}},
  \bibnamefont{et~al.}, \bibinfo{journal}{\apjs}
  \textbf{\bibinfo{volume}{199}}, \bibinfo{eid}{25} (\bibinfo{year}{2012}),
  \eprint{1106.3328}.

\bibitem[{\citenamefont{{Umetsu} et~al.}(2016)\citenamefont{{Umetsu}, {Zitrin},
  {Gruen}, {Merten}, {Donahue}, and {Postman}}}]{Umetsu16}
\bibinfo{author}{\bibfnamefont{K.}~\bibnamefont{{Umetsu}}},
  \bibinfo{author}{\bibfnamefont{A.}~\bibnamefont{{Zitrin}}},
  \bibinfo{author}{\bibfnamefont{D.}~\bibnamefont{{Gruen}}},
  \bibinfo{author}{\bibfnamefont{J.}~\bibnamefont{{Merten}}},
  \bibinfo{author}{\bibfnamefont{M.}~\bibnamefont{{Donahue}}},
  \bibnamefont{and}
  \bibinfo{author}{\bibfnamefont{M.}~\bibnamefont{{Postman}}},
  \bibinfo{journal}{\apj} \textbf{\bibinfo{volume}{821}}, \bibinfo{eid}{116}
  (\bibinfo{year}{2016}), \eprint{1507.04385}.

\bibitem[{\citenamefont{{Donahue} et~al.}(2014)\citenamefont{{Donahue}, {Voit},
  {Mahdavi}, {Umetsu}, {Ettori}, {Merten}, {Postman}, {Hoffer}, {Baldi}, {Coe}
  et~al.}}]{Donahue14}
\bibinfo{author}{\bibfnamefont{M.}~\bibnamefont{{Donahue}}},
  \bibinfo{author}{\bibfnamefont{G.~M.} \bibnamefont{{Voit}}},
  \bibinfo{author}{\bibfnamefont{A.}~\bibnamefont{{Mahdavi}}},
  \bibinfo{author}{\bibfnamefont{K.}~\bibnamefont{{Umetsu}}},
  \bibinfo{author}{\bibfnamefont{S.}~\bibnamefont{{Ettori}}},
  \bibinfo{author}{\bibfnamefont{J.}~\bibnamefont{{Merten}}},
  \bibinfo{author}{\bibfnamefont{M.}~\bibnamefont{{Postman}}},
  \bibinfo{author}{\bibfnamefont{A.}~\bibnamefont{{Hoffer}}},
  \bibinfo{author}{\bibfnamefont{A.}~\bibnamefont{{Baldi}}},
  \bibinfo{author}{\bibfnamefont{D.}~\bibnamefont{{Coe}}},
  \bibnamefont{et~al.}, \bibinfo{journal}{\apj} \textbf{\bibinfo{volume}{794}},
  \bibinfo{eid}{136} (\bibinfo{year}{2014}), \eprint{1405.7876}.

\bibitem[{\citenamefont{{Navarro} et~al.}(1997)\citenamefont{{Navarro},
  {Frenk}, and {White}}}]{NFW}
\bibinfo{author}{\bibfnamefont{J.~F.} \bibnamefont{{Navarro}}},
  \bibinfo{author}{\bibfnamefont{C.~S.} \bibnamefont{{Frenk}}},
  \bibnamefont{and} \bibinfo{author}{\bibfnamefont{S.~D.~M.}
  \bibnamefont{{White}}}, \bibinfo{journal}{\apj}
  \textbf{\bibinfo{volume}{490}}, \bibinfo{pages}{493} (\bibinfo{year}{1997}),
  \eprint{astro-ph/9611107}.

\bibitem[{\citenamefont{{Chiu} et~al.}(2018)\citenamefont{{Chiu}, {Mohr},
  {McDonald}, {Bocquet}, {Desai}, {Klein}, {Israel}, {Ashby}, {Stanford},
  {Benson} et~al.}}]{Chiu18}
\bibinfo{author}{\bibfnamefont{I.}~\bibnamefont{{Chiu}}},
  \bibinfo{author}{\bibfnamefont{J.~J.} \bibnamefont{{Mohr}}},
  \bibinfo{author}{\bibfnamefont{M.}~\bibnamefont{{McDonald}}},
  \bibinfo{author}{\bibfnamefont{S.}~\bibnamefont{{Bocquet}}},
  \bibinfo{author}{\bibfnamefont{S.}~\bibnamefont{{Desai}}},
  \bibinfo{author}{\bibfnamefont{M.}~\bibnamefont{{Klein}}},
  \bibinfo{author}{\bibfnamefont{H.}~\bibnamefont{{Israel}}},
  \bibinfo{author}{\bibfnamefont{M.~L.~N.} \bibnamefont{{Ashby}}},
  \bibinfo{author}{\bibfnamefont{A.}~\bibnamefont{{Stanford}}},
  \bibinfo{author}{\bibfnamefont{B.~A.} \bibnamefont{{Benson}}},
  \bibnamefont{et~al.}, \bibinfo{journal}{\mnras}
  \textbf{\bibinfo{volume}{478}}, \bibinfo{pages}{3072} (\bibinfo{year}{2018}),
  \eprint{1711.00917}.

\bibitem[{\citenamefont{{Hernquist}}(1990)}]{Hernquist}
\bibinfo{author}{\bibfnamefont{L.}~\bibnamefont{{Hernquist}}},
  \bibinfo{journal}{\apj} \textbf{\bibinfo{volume}{356}}, \bibinfo{pages}{359}
  (\bibinfo{year}{1990}).

\bibitem[{\citenamefont{{Cooke} et~al.}(2016)\citenamefont{{Cooke}, {O'Dea},
  {Baum}, {Tremblay}, {Cox}, and {Gladders}}}]{Cooke}
\bibinfo{author}{\bibfnamefont{K.~C.} \bibnamefont{{Cooke}}},
  \bibinfo{author}{\bibfnamefont{C.~P.} \bibnamefont{{O'Dea}}},
  \bibinfo{author}{\bibfnamefont{S.~A.} \bibnamefont{{Baum}}},
  \bibinfo{author}{\bibfnamefont{G.~R.} \bibnamefont{{Tremblay}}},
  \bibinfo{author}{\bibfnamefont{I.~G.} \bibnamefont{{Cox}}}, \bibnamefont{and}
  \bibinfo{author}{\bibfnamefont{M.}~\bibnamefont{{Gladders}}},
  \bibinfo{journal}{\apj} \textbf{\bibinfo{volume}{833}}, \bibinfo{eid}{224}
  (\bibinfo{year}{2016}), \eprint{1610.05310}.

\bibitem[{\citenamefont{{Olamaie} et~al.}(2012)\citenamefont{{Olamaie},
  {Hobson}, and {Grainge}}}]{Olamaie}
\bibinfo{author}{\bibfnamefont{M.}~\bibnamefont{{Olamaie}}},
  \bibinfo{author}{\bibfnamefont{M.~P.} \bibnamefont{{Hobson}}},
  \bibnamefont{and} \bibinfo{author}{\bibfnamefont{K.~J.~B.}
  \bibnamefont{{Grainge}}}, \bibinfo{journal}{\mnras}
  \textbf{\bibinfo{volume}{423}}, \bibinfo{pages}{1534} (\bibinfo{year}{2012}),
  \eprint{1109.2834}.

\bibitem[{\citenamefont{{Tian} et~al.}(2020{\natexlab{b}})\citenamefont{{Tian},
  {Yu}, {Li}, {McGaugh}, and {Ko}}}]{Ko20}
\bibinfo{author}{\bibfnamefont{Y.}~\bibnamefont{{Tian}}},
  \bibinfo{author}{\bibfnamefont{P.-C.} \bibnamefont{{Yu}}},
  \bibinfo{author}{\bibfnamefont{P.}~\bibnamefont{{Li}}},
  \bibinfo{author}{\bibfnamefont{S.~S.} \bibnamefont{{McGaugh}}},
  \bibnamefont{and} \bibinfo{author}{\bibfnamefont{C.-M.} \bibnamefont{{Ko}}},
  \bibinfo{journal}{arXiv e-prints} \bibinfo{eid}{arXiv:2010.00992}
  (\bibinfo{year}{2020}{\natexlab{b}}), \eprint{2010.00992}.

\bibitem[{\citenamefont{{Vikhlinin} et~al.}(2005)\citenamefont{{Vikhlinin},
  {Markevitch}, {Murray}, {Jones}, {Forman}, and {Van
  Speybroeck}}}]{Vikhlinin05}
\bibinfo{author}{\bibfnamefont{A.}~\bibnamefont{{Vikhlinin}}},
  \bibinfo{author}{\bibfnamefont{M.}~\bibnamefont{{Markevitch}}},
  \bibinfo{author}{\bibfnamefont{S.~S.} \bibnamefont{{Murray}}},
  \bibinfo{author}{\bibfnamefont{C.}~\bibnamefont{{Jones}}},
  \bibinfo{author}{\bibfnamefont{W.}~\bibnamefont{{Forman}}}, \bibnamefont{and}
  \bibinfo{author}{\bibfnamefont{L.}~\bibnamefont{{Van Speybroeck}}},
  \bibinfo{journal}{\apj} \textbf{\bibinfo{volume}{628}}, \bibinfo{pages}{655}
  (\bibinfo{year}{2005}), \eprint{astro-ph/0412306}.

\bibitem[{\citenamefont{{Vikhlinin} et~al.}(2006)\citenamefont{{Vikhlinin},
  {Kravtsov}, {Forman}, {Jones}, {Markevitch}, {Murray}, and {Van
  Speybroeck}}}]{Vikhlinin06}
\bibinfo{author}{\bibfnamefont{A.}~\bibnamefont{{Vikhlinin}}},
  \bibinfo{author}{\bibfnamefont{A.}~\bibnamefont{{Kravtsov}}},
  \bibinfo{author}{\bibfnamefont{W.}~\bibnamefont{{Forman}}},
  \bibinfo{author}{\bibfnamefont{C.}~\bibnamefont{{Jones}}},
  \bibinfo{author}{\bibfnamefont{M.}~\bibnamefont{{Markevitch}}},
  \bibinfo{author}{\bibfnamefont{S.~S.} \bibnamefont{{Murray}}},
  \bibnamefont{and} \bibinfo{author}{\bibfnamefont{L.}~\bibnamefont{{Van
  Speybroeck}}}, \bibinfo{journal}{\apj} \textbf{\bibinfo{volume}{640}},
  \bibinfo{pages}{691} (\bibinfo{year}{2006}), \eprint{astro-ph/0507092}.

\bibitem[{\citenamefont{{Rahvar} and {Mashhoon}}(2014)}]{Rahvar}
\bibinfo{author}{\bibfnamefont{S.}~\bibnamefont{{Rahvar}}} \bibnamefont{and}
  \bibinfo{author}{\bibfnamefont{B.}~\bibnamefont{{Mashhoon}}},
  \bibinfo{journal}{\prd} \textbf{\bibinfo{volume}{89}}, \bibinfo{eid}{104011}
  (\bibinfo{year}{2014}), \eprint{1401.4819}.

\bibitem[{\citenamefont{{Hodson} and {Zhao}}(2017)}]{Hodson17}
\bibinfo{author}{\bibfnamefont{A.~O.} \bibnamefont{{Hodson}}} \bibnamefont{and}
  \bibinfo{author}{\bibfnamefont{H.}~\bibnamefont{{Zhao}}},
  \bibinfo{journal}{\aap} \textbf{\bibinfo{volume}{598}}, \bibinfo{eid}{A127}
  (\bibinfo{year}{2017}), \eprint{1701.03369}.

\bibitem[{\citenamefont{{Hodson} et~al.}(2017)\citenamefont{{Hodson}, {Zhao},
  {Khoury}, and {Famaey}}}]{Khoury17}
\bibinfo{author}{\bibfnamefont{A.~O.} \bibnamefont{{Hodson}}},
  \bibinfo{author}{\bibfnamefont{H.}~\bibnamefont{{Zhao}}},
  \bibinfo{author}{\bibfnamefont{J.}~\bibnamefont{{Khoury}}}, \bibnamefont{and}
  \bibinfo{author}{\bibfnamefont{B.}~\bibnamefont{{Famaey}}},
  \bibinfo{journal}{\aap} \textbf{\bibinfo{volume}{607}}, \bibinfo{eid}{A108}
  (\bibinfo{year}{2017}), \eprint{1611.05876}.

\bibitem[{\citenamefont{{Edmonds} et~al.}(2018)\citenamefont{{Edmonds},
  {Farrah}, {Minic}, {Ng}, and {Takeuchi}}}]{Ng}
\bibinfo{author}{\bibfnamefont{D.}~\bibnamefont{{Edmonds}}},
  \bibinfo{author}{\bibfnamefont{D.}~\bibnamefont{{Farrah}}},
  \bibinfo{author}{\bibfnamefont{D.}~\bibnamefont{{Minic}}},
  \bibinfo{author}{\bibfnamefont{Y.~J.} \bibnamefont{{Ng}}}, \bibnamefont{and}
  \bibinfo{author}{\bibfnamefont{T.}~\bibnamefont{{Takeuchi}}},
  \bibinfo{journal}{International Journal of Modern Physics D}
  \textbf{\bibinfo{volume}{27}}, \bibinfo{eid}{1830001-296}
  (\bibinfo{year}{2018}), \eprint{1709.04388}.

\bibitem[{\citenamefont{{Edmonds} et~al.}(2017)\citenamefont{{Edmonds},
  {Farrah}, {Ho}, {Minic}, {Ng}, and {Takeuchi}}}]{Edmonds}
\bibinfo{author}{\bibfnamefont{D.}~\bibnamefont{{Edmonds}}},
  \bibinfo{author}{\bibfnamefont{D.}~\bibnamefont{{Farrah}}},
  \bibinfo{author}{\bibfnamefont{C.~M.} \bibnamefont{{Ho}}},
  \bibinfo{author}{\bibfnamefont{D.}~\bibnamefont{{Minic}}},
  \bibinfo{author}{\bibfnamefont{Y.~J.} \bibnamefont{{Ng}}}, \bibnamefont{and}
  \bibinfo{author}{\bibfnamefont{T.}~\bibnamefont{{Takeuchi}}},
  \bibinfo{journal}{International Journal of Modern Physics A}
  \textbf{\bibinfo{volume}{32}}, \bibinfo{eid}{1750108} (\bibinfo{year}{2017}),
  \eprint{1601.00662}.

\bibitem[{\citenamefont{{Bernal} et~al.}(2017)\citenamefont{{Bernal}, {Robles},
  and {Matos}}}]{Bernal}
\bibinfo{author}{\bibfnamefont{T.}~\bibnamefont{{Bernal}}},
  \bibinfo{author}{\bibfnamefont{V.~H.} \bibnamefont{{Robles}}},
  \bibnamefont{and} \bibinfo{author}{\bibfnamefont{T.}~\bibnamefont{{Matos}}},
  \bibinfo{journal}{\mnras} \textbf{\bibinfo{volume}{468}},
  \bibinfo{pages}{3135} (\bibinfo{year}{2017}), \eprint{1609.08644}.

\bibitem[{\citenamefont{{Gupta} and {Desai}}(2019)}]{Gupta1}
\bibinfo{author}{\bibfnamefont{S.}~\bibnamefont{{Gupta}}} \bibnamefont{and}
  \bibinfo{author}{\bibfnamefont{S.}~\bibnamefont{{Desai}}},
  \bibinfo{journal}{Classical and Quantum Gravity}
  \textbf{\bibinfo{volume}{36}}, \bibinfo{eid}{105001} (\bibinfo{year}{2019}),
  \eprint{1811.09378}.

\bibitem[{\citenamefont{{Gupta} and {Desai}}(2020)}]{Gupta2}
\bibinfo{author}{\bibfnamefont{S.}~\bibnamefont{{Gupta}}} \bibnamefont{and}
  \bibinfo{author}{\bibfnamefont{S.}~\bibnamefont{{Desai}}},
  \bibinfo{journal}{Physics of the Dark Universe}
  \textbf{\bibinfo{volume}{28}}, \bibinfo{eid}{100499} (\bibinfo{year}{2020}),
  \eprint{1909.07408}.

\bibitem[{\citenamefont{{Gopika} and {Desai}}(2020)}]{Gopika}
\bibinfo{author}{\bibfnamefont{K.}~\bibnamefont{{Gopika}}} \bibnamefont{and}
  \bibinfo{author}{\bibfnamefont{S.}~\bibnamefont{{Desai}}},
  \bibinfo{journal}{Physics of the Dark Universe}
  \textbf{\bibinfo{volume}{30}}, \bibinfo{eid}{100707} (\bibinfo{year}{2020}),
  \eprint{2006.12320}.

\bibitem[{\citenamefont{{Tian} and {Ko}}(2019)}]{Tian19}
\bibinfo{author}{\bibfnamefont{Y.}~\bibnamefont{{Tian}}} \bibnamefont{and}
  \bibinfo{author}{\bibfnamefont{C.-M.} \bibnamefont{{Ko}}},
  \bibinfo{journal}{\mnras} \textbf{\bibinfo{volume}{488}},
  \bibinfo{pages}{L41} (\bibinfo{year}{2019}), \eprint{1907.04501}.

\bibitem[{\citenamefont{{Biffi} et~al.}(2016)\citenamefont{{Biffi}, {Borgani},
  {Murante}, {Rasia}, {Planelles}, {Granato}, {Ragone-Figueroa}, {Beck},
  {Gaspari}, and {Dolag}}}]{Biffi16}
\bibinfo{author}{\bibfnamefont{V.}~\bibnamefont{{Biffi}}},
  \bibinfo{author}{\bibfnamefont{S.}~\bibnamefont{{Borgani}}},
  \bibinfo{author}{\bibfnamefont{G.}~\bibnamefont{{Murante}}},
  \bibinfo{author}{\bibfnamefont{E.}~\bibnamefont{{Rasia}}},
  \bibinfo{author}{\bibfnamefont{S.}~\bibnamefont{{Planelles}}},
  \bibinfo{author}{\bibfnamefont{G.~L.} \bibnamefont{{Granato}}},
  \bibinfo{author}{\bibfnamefont{C.}~\bibnamefont{{Ragone-Figueroa}}},
  \bibinfo{author}{\bibfnamefont{A.~M.} \bibnamefont{{Beck}}},
  \bibinfo{author}{\bibfnamefont{M.}~\bibnamefont{{Gaspari}}},
  \bibnamefont{and} \bibinfo{author}{\bibfnamefont{K.}~\bibnamefont{{Dolag}}},
  \bibinfo{journal}{\apj} \textbf{\bibinfo{volume}{827}}, \bibinfo{eid}{112}
  (\bibinfo{year}{2016}), \eprint{1606.02293}.

\bibitem[{\citenamefont{{Lin} et~al.}(2012)\citenamefont{{Lin}, {Stanford},
  {Eisenhardt}, {Vikhlinin}, {Maughan}, and {Kravtsov}}}]{ytlin}
\bibinfo{author}{\bibfnamefont{Y.-T.} \bibnamefont{{Lin}}},
  \bibinfo{author}{\bibfnamefont{S.~A.} \bibnamefont{{Stanford}}},
  \bibinfo{author}{\bibfnamefont{P.~R.~M.} \bibnamefont{{Eisenhardt}}},
  \bibinfo{author}{\bibfnamefont{A.}~\bibnamefont{{Vikhlinin}}},
  \bibinfo{author}{\bibfnamefont{B.~J.} \bibnamefont{{Maughan}}},
  \bibnamefont{and}
  \bibinfo{author}{\bibfnamefont{A.}~\bibnamefont{{Kravtsov}}},
  \bibinfo{journal}{\apjl} \textbf{\bibinfo{volume}{745}}, \bibinfo{eid}{L3}
  (\bibinfo{year}{2012}), \eprint{1112.1705}.

\bibitem[{\citenamefont{{Vikhlinin} et~al.}(2009)\citenamefont{{Vikhlinin},
  {Burenin}, {Ebeling}, {Forman}, {Hornstrup}, {Jones}, {Kravtsov}, {Murray},
  {Nagai}, {Quintana} et~al.}}]{Vikhlinin09}
\bibinfo{author}{\bibfnamefont{A.}~\bibnamefont{{Vikhlinin}}},
  \bibinfo{author}{\bibfnamefont{R.~A.} \bibnamefont{{Burenin}}},
  \bibinfo{author}{\bibfnamefont{H.}~\bibnamefont{{Ebeling}}},
  \bibinfo{author}{\bibfnamefont{W.~R.} \bibnamefont{{Forman}}},
  \bibinfo{author}{\bibfnamefont{A.}~\bibnamefont{{Hornstrup}}},
  \bibinfo{author}{\bibfnamefont{C.}~\bibnamefont{{Jones}}},
  \bibinfo{author}{\bibfnamefont{A.~V.} \bibnamefont{{Kravtsov}}},
  \bibinfo{author}{\bibfnamefont{S.~S.} \bibnamefont{{Murray}}},
  \bibinfo{author}{\bibfnamefont{D.}~\bibnamefont{{Nagai}}},
  \bibinfo{author}{\bibfnamefont{H.}~\bibnamefont{{Quintana}}},
  \bibnamefont{et~al.}, \bibinfo{journal}{\apj} \textbf{\bibinfo{volume}{692}},
  \bibinfo{pages}{1033} (\bibinfo{year}{2009}), \eprint{0805.2207}.

\bibitem[{\citenamefont{Foreman-Mackey
  et~al.}(2013)\citenamefont{Foreman-Mackey, Hogg, Lang, and Goodman}}]{emcee}
\bibinfo{author}{\bibfnamefont{D.}~\bibnamefont{Foreman-Mackey}},
  \bibinfo{author}{\bibfnamefont{D.~W.} \bibnamefont{Hogg}},
  \bibinfo{author}{\bibfnamefont{D.}~\bibnamefont{Lang}}, \bibnamefont{and}
  \bibinfo{author}{\bibfnamefont{J.}~\bibnamefont{Goodman}},
  \bibinfo{journal}{Publ. Astron. Soc. Pac.} \textbf{\bibinfo{volume}{125}},
  \bibinfo{pages}{306} (\bibinfo{year}{2013}), \eprint{1202.3665}.

\bibitem[{\citenamefont{{Astropy Collaboration}
  et~al.}(2018)\citenamefont{{Astropy Collaboration}, {Price-Whelan},
  {Sip{\H{o}}cz}, {G{\"u}nther}, {Lim}, {Crawford}, {Conseil}, {Shupe},
  {Craig}, {Dencheva} et~al.}}]{astropy}
\bibinfo{author}{\bibnamefont{{Astropy Collaboration}}},
  \bibinfo{author}{\bibfnamefont{A.~M.} \bibnamefont{{Price-Whelan}}},
  \bibinfo{author}{\bibfnamefont{B.~M.} \bibnamefont{{Sip{\H{o}}cz}}},
  \bibinfo{author}{\bibfnamefont{H.~M.} \bibnamefont{{G{\"u}nther}}},
  \bibinfo{author}{\bibfnamefont{P.~L.} \bibnamefont{{Lim}}},
  \bibinfo{author}{\bibfnamefont{S.~M.} \bibnamefont{{Crawford}}},
  \bibinfo{author}{\bibfnamefont{S.}~\bibnamefont{{Conseil}}},
  \bibinfo{author}{\bibfnamefont{D.~L.} \bibnamefont{{Shupe}}},
  \bibinfo{author}{\bibfnamefont{M.~W.} \bibnamefont{{Craig}}},
  \bibinfo{author}{\bibfnamefont{N.}~\bibnamefont{{Dencheva}}},
  \bibnamefont{et~al.}, \bibinfo{journal}{\aj} \textbf{\bibinfo{volume}{156}},
  \bibinfo{eid}{123} (\bibinfo{year}{2018}), \eprint{1801.02634}.

\bibitem[{\citenamefont{{Planck Collaboration}
  et~al.}(2014)\citenamefont{{Planck Collaboration}, {Ade}, {Aghanim},
  {Armitage-Caplan}, {Arnaud}, {Ashdown}, {Atrio-Barand ela}, {Aumont},
  {Aussel}, {Baccigalupi} et~al.}}]{planck_2014}
\bibinfo{author}{\bibnamefont{{Planck Collaboration}}},
  \bibinfo{author}{\bibfnamefont{P.~A.~R.} \bibnamefont{{Ade}}},
  \bibinfo{author}{\bibfnamefont{N.}~\bibnamefont{{Aghanim}}},
  \bibinfo{author}{\bibfnamefont{C.}~\bibnamefont{{Armitage-Caplan}}},
  \bibinfo{author}{\bibfnamefont{M.}~\bibnamefont{{Arnaud}}},
  \bibinfo{author}{\bibfnamefont{M.}~\bibnamefont{{Ashdown}}},
  \bibinfo{author}{\bibfnamefont{F.}~\bibnamefont{{Atrio-Barand ela}}},
  \bibinfo{author}{\bibfnamefont{J.}~\bibnamefont{{Aumont}}},
  \bibinfo{author}{\bibfnamefont{H.}~\bibnamefont{{Aussel}}},
  \bibinfo{author}{\bibfnamefont{C.}~\bibnamefont{{Baccigalupi}}},
  \bibnamefont{et~al.}, \bibinfo{journal}{\aap} \textbf{\bibinfo{volume}{571}},
  \bibinfo{eid}{A29} (\bibinfo{year}{2014}), \eprint{1303.5089}.

\bibitem[{\citenamefont{{Ghirardini} et~al.}(2019)\citenamefont{{Ghirardini},
  {Eckert}, {Ettori}, {Pointecouteau}, {Molendi}, {Gaspari}, {Rossetti}, {De
  Grandi}, {Roncarelli}, {Bourdin} et~al.}}]{xcop_thermo}
\bibinfo{author}{\bibfnamefont{V.}~\bibnamefont{{Ghirardini}}},
  \bibinfo{author}{\bibfnamefont{D.}~\bibnamefont{{Eckert}}},
  \bibinfo{author}{\bibfnamefont{S.}~\bibnamefont{{Ettori}}},
  \bibinfo{author}{\bibfnamefont{E.}~\bibnamefont{{Pointecouteau}}},
  \bibinfo{author}{\bibfnamefont{S.}~\bibnamefont{{Molendi}}},
  \bibinfo{author}{\bibfnamefont{M.}~\bibnamefont{{Gaspari}}},
  \bibinfo{author}{\bibfnamefont{M.}~\bibnamefont{{Rossetti}}},
  \bibinfo{author}{\bibfnamefont{S.}~\bibnamefont{{De Grandi}}},
  \bibinfo{author}{\bibfnamefont{M.}~\bibnamefont{{Roncarelli}}},
  \bibinfo{author}{\bibfnamefont{H.}~\bibnamefont{{Bourdin}}},
  \bibnamefont{et~al.}, \bibinfo{journal}{\aap} \textbf{\bibinfo{volume}{621}},
  \bibinfo{eid}{A41} (\bibinfo{year}{2019}), \eprint{1805.00042}.

\bibitem[{\citenamefont{{Eckert} et~al.}(2017)\citenamefont{{Eckert}, {Ettori},
  {Pointecouteau}, {Molendi}, {Paltani}, and {Tchernin}}}]{XCOPmain}
\bibinfo{author}{\bibfnamefont{D.}~\bibnamefont{{Eckert}}},
  \bibinfo{author}{\bibfnamefont{S.}~\bibnamefont{{Ettori}}},
  \bibinfo{author}{\bibfnamefont{E.}~\bibnamefont{{Pointecouteau}}},
  \bibinfo{author}{\bibfnamefont{S.}~\bibnamefont{{Molendi}}},
  \bibinfo{author}{\bibfnamefont{S.}~\bibnamefont{{Paltani}}},
  \bibnamefont{and}
  \bibinfo{author}{\bibfnamefont{C.}~\bibnamefont{{Tchernin}}},
  \bibinfo{journal}{Astronomische Nachrichten} \textbf{\bibinfo{volume}{338}},
  \bibinfo{pages}{293} (\bibinfo{year}{2017}), \eprint{1611.05051}.

\bibitem[{\citenamefont{{Ettori} et~al.}(2019)\citenamefont{{Ettori},
  {Ghirardini}, {Eckert}, {Pointecouteau}, {Gastaldello}, {Sereno}, {Gaspari},
  {Ghizzardi}, {Roncarelli}, and {Rossetti}}}]{XCOP1}
\bibinfo{author}{\bibfnamefont{S.}~\bibnamefont{{Ettori}}},
  \bibinfo{author}{\bibfnamefont{V.}~\bibnamefont{{Ghirardini}}},
  \bibinfo{author}{\bibfnamefont{D.}~\bibnamefont{{Eckert}}},
  \bibinfo{author}{\bibfnamefont{E.}~\bibnamefont{{Pointecouteau}}},
  \bibinfo{author}{\bibfnamefont{F.}~\bibnamefont{{Gastaldello}}},
  \bibinfo{author}{\bibfnamefont{M.}~\bibnamefont{{Sereno}}},
  \bibinfo{author}{\bibfnamefont{M.}~\bibnamefont{{Gaspari}}},
  \bibinfo{author}{\bibfnamefont{S.}~\bibnamefont{{Ghizzardi}}},
  \bibinfo{author}{\bibfnamefont{M.}~\bibnamefont{{Roncarelli}}},
  \bibnamefont{and}
  \bibinfo{author}{\bibfnamefont{M.}~\bibnamefont{{Rossetti}}},
  \bibinfo{journal}{\aap} \textbf{\bibinfo{volume}{621}}, \bibinfo{eid}{A39}
  (\bibinfo{year}{2019}), \eprint{1805.00035}.

\bibitem[{\citenamefont{{Ghirardini} et~al.}(2018)\citenamefont{{Ghirardini},
  {Ettori}, {Eckert}, {Molendi}, {Gastaldello}, {Pointecouteau}, {Hurier}, and
  {Bourdin}}}]{xcop_thermoa2319}
\bibinfo{author}{\bibfnamefont{V.}~\bibnamefont{{Ghirardini}}},
  \bibinfo{author}{\bibfnamefont{S.}~\bibnamefont{{Ettori}}},
  \bibinfo{author}{\bibfnamefont{D.}~\bibnamefont{{Eckert}}},
  \bibinfo{author}{\bibfnamefont{S.}~\bibnamefont{{Molendi}}},
  \bibinfo{author}{\bibfnamefont{F.}~\bibnamefont{{Gastaldello}}},
  \bibinfo{author}{\bibfnamefont{E.}~\bibnamefont{{Pointecouteau}}},
  \bibinfo{author}{\bibfnamefont{G.}~\bibnamefont{{Hurier}}}, \bibnamefont{and}
  \bibinfo{author}{\bibfnamefont{H.}~\bibnamefont{{Bourdin}}},
  \bibinfo{journal}{\aap} \textbf{\bibinfo{volume}{614}}, \bibinfo{eid}{A7}
  (\bibinfo{year}{2018}), \eprint{1708.02954}.

\bibitem[{\citenamefont{{Ettori} et~al.}(2020)\citenamefont{{Ettori},
  {Ghirardini}, and {Eckert}}}]{xcop_He}
\bibinfo{author}{\bibfnamefont{S.}~\bibnamefont{{Ettori}}},
  \bibinfo{author}{\bibfnamefont{V.}~\bibnamefont{{Ghirardini}}},
  \bibnamefont{and} \bibinfo{author}{\bibfnamefont{D.}~\bibnamefont{{Eckert}}},
  \bibinfo{journal}{Astronomische Nachrichten} \textbf{\bibinfo{volume}{341}},
  \bibinfo{pages}{210} (\bibinfo{year}{2020}), \eprint{1911.05039}.

\bibitem[{\citenamefont{{Ghizzardi}
  et~al.}(2020{\natexlab{a}})\citenamefont{{Ghizzardi}, {Molendi}, {van der
  Burg}, {De Grandi}, {Bartalucci}, {Gastaldello}, {Rossetti}, {Biffi},
  {Borgani}, {Eckert} et~al.}}]{Ghizzardi}
\bibinfo{author}{\bibfnamefont{S.}~\bibnamefont{{Ghizzardi}}},
  \bibinfo{author}{\bibfnamefont{S.}~\bibnamefont{{Molendi}}},
  \bibinfo{author}{\bibfnamefont{R.}~\bibnamefont{{van der Burg}}},
  \bibinfo{author}{\bibfnamefont{S.}~\bibnamefont{{De Grandi}}},
  \bibinfo{author}{\bibfnamefont{I.}~\bibnamefont{{Bartalucci}}},
  \bibinfo{author}{\bibfnamefont{F.}~\bibnamefont{{Gastaldello}}},
  \bibinfo{author}{\bibfnamefont{M.}~\bibnamefont{{Rossetti}}},
  \bibinfo{author}{\bibfnamefont{V.}~\bibnamefont{{Biffi}}},
  \bibinfo{author}{\bibfnamefont{S.}~\bibnamefont{{Borgani}}},
  \bibinfo{author}{\bibfnamefont{D.}~\bibnamefont{{Eckert}}},
  \bibnamefont{et~al.}, \bibinfo{journal}{arXiv e-prints}
  \bibinfo{eid}{arXiv:2007.01084} (\bibinfo{year}{2020}{\natexlab{a}}),
  \eprint{2007.01084}.

\bibitem[{\citenamefont{{Sunyaev} and {Zeldovich}}(1970)}]{SZ}
\bibinfo{author}{\bibfnamefont{R.~A.} \bibnamefont{{Sunyaev}}}
  \bibnamefont{and} \bibinfo{author}{\bibfnamefont{Y.~B.}
  \bibnamefont{{Zeldovich}}}, \bibinfo{journal}{\apss}
  \textbf{\bibinfo{volume}{7}}, \bibinfo{pages}{3} (\bibinfo{year}{1970}).

\bibitem[{\citenamefont{{Planck Collaboration}
  et~al.}(2016)\citenamefont{{Planck Collaboration}, {Ade}, {Aghanim},
  {Arnaud}, {Ashdown}, {Aumont}, {Baccigalupi}, {Banday}, {Barreiro}, {Barrena}
  et~al.}}]{Planck_2015}
\bibinfo{author}{\bibnamefont{{Planck Collaboration}}},
  \bibinfo{author}{\bibfnamefont{P.~A.~R.} \bibnamefont{{Ade}}},
  \bibinfo{author}{\bibfnamefont{N.}~\bibnamefont{{Aghanim}}},
  \bibinfo{author}{\bibfnamefont{M.}~\bibnamefont{{Arnaud}}},
  \bibinfo{author}{\bibfnamefont{M.}~\bibnamefont{{Ashdown}}},
  \bibinfo{author}{\bibfnamefont{J.}~\bibnamefont{{Aumont}}},
  \bibinfo{author}{\bibfnamefont{C.}~\bibnamefont{{Baccigalupi}}},
  \bibinfo{author}{\bibfnamefont{A.~J.} \bibnamefont{{Banday}}},
  \bibinfo{author}{\bibfnamefont{R.~B.} \bibnamefont{{Barreiro}}},
  \bibinfo{author}{\bibfnamefont{R.}~\bibnamefont{{Barrena}}},
  \bibnamefont{et~al.}, \bibinfo{journal}{\aap} \textbf{\bibinfo{volume}{594}},
  \bibinfo{eid}{A27} (\bibinfo{year}{2016}), \eprint{1502.01598}.

\bibitem[{\citenamefont{{Eckert} et~al.}(2019)\citenamefont{{Eckert},
  {Ghirardini}, {Ettori}, {Rasia}, {Biffi}, {Pointecouteau}, {Rossetti},
  {Molendi}, {Vazza}, {Gastaldello} et~al.}}]{xcop_fgas}
\bibinfo{author}{\bibfnamefont{D.}~\bibnamefont{{Eckert}}},
  \bibinfo{author}{\bibfnamefont{V.}~\bibnamefont{{Ghirardini}}},
  \bibinfo{author}{\bibfnamefont{S.}~\bibnamefont{{Ettori}}},
  \bibinfo{author}{\bibfnamefont{E.}~\bibnamefont{{Rasia}}},
  \bibinfo{author}{\bibfnamefont{V.}~\bibnamefont{{Biffi}}},
  \bibinfo{author}{\bibfnamefont{E.}~\bibnamefont{{Pointecouteau}}},
  \bibinfo{author}{\bibfnamefont{M.}~\bibnamefont{{Rossetti}}},
  \bibinfo{author}{\bibfnamefont{S.}~\bibnamefont{{Molendi}}},
  \bibinfo{author}{\bibfnamefont{F.}~\bibnamefont{{Vazza}}},
  \bibinfo{author}{\bibfnamefont{F.}~\bibnamefont{{Gastaldello}}},
  \bibnamefont{et~al.}, \bibinfo{journal}{\aap} \textbf{\bibinfo{volume}{621}},
  \bibinfo{eid}{A40} (\bibinfo{year}{2019}), \eprint{1805.00034}.

\bibitem[{\citenamefont{{Ghizzardi}
  et~al.}(2020{\natexlab{b}})\citenamefont{{Ghizzardi}, {Molendi}, {van der
  Burg}, {De Grandi}, {Bartalucci}, {Gastaldello}, {Rossetti}, {Biffi},
  {Borgani}, {Eckert} et~al.}}]{xcop_mstar}
\bibinfo{author}{\bibfnamefont{S.}~\bibnamefont{{Ghizzardi}}},
  \bibinfo{author}{\bibfnamefont{S.}~\bibnamefont{{Molendi}}},
  \bibinfo{author}{\bibfnamefont{R.}~\bibnamefont{{van der Burg}}},
  \bibinfo{author}{\bibfnamefont{S.}~\bibnamefont{{De Grandi}}},
  \bibinfo{author}{\bibfnamefont{I.}~\bibnamefont{{Bartalucci}}},
  \bibinfo{author}{\bibfnamefont{F.}~\bibnamefont{{Gastaldello}}},
  \bibinfo{author}{\bibfnamefont{M.}~\bibnamefont{{Rossetti}}},
  \bibinfo{author}{\bibfnamefont{V.}~\bibnamefont{{Biffi}}},
  \bibinfo{author}{\bibfnamefont{S.}~\bibnamefont{{Borgani}}},
  \bibinfo{author}{\bibfnamefont{D.}~\bibnamefont{{Eckert}}},
  \bibnamefont{et~al.}, \bibinfo{journal}{arXiv e-prints}
  \bibinfo{eid}{arXiv:2007.01084} (\bibinfo{year}{2020}{\natexlab{b}}),
  \eprint{2007.01084}.

\bibitem[{\citenamefont{{Predehl} et~al.}(2020)\citenamefont{{Predehl},
  {Andritschke}, {Arefiev}, {Babyshkin}, {Batanov}, {Becker}, {B{\"o}hringer},
  {Bogomolov}, {Boller}, {Borm} et~al.}}]{erosita}
\bibinfo{author}{\bibfnamefont{P.}~\bibnamefont{{Predehl}}},
  \bibinfo{author}{\bibfnamefont{R.}~\bibnamefont{{Andritschke}}},
  \bibinfo{author}{\bibfnamefont{V.}~\bibnamefont{{Arefiev}}},
  \bibinfo{author}{\bibfnamefont{V.}~\bibnamefont{{Babyshkin}}},
  \bibinfo{author}{\bibfnamefont{O.}~\bibnamefont{{Batanov}}},
  \bibinfo{author}{\bibfnamefont{W.}~\bibnamefont{{Becker}}},
  \bibinfo{author}{\bibfnamefont{H.}~\bibnamefont{{B{\"o}hringer}}},
  \bibinfo{author}{\bibfnamefont{A.}~\bibnamefont{{Bogomolov}}},
  \bibinfo{author}{\bibfnamefont{T.}~\bibnamefont{{Boller}}},
  \bibinfo{author}{\bibfnamefont{K.}~\bibnamefont{{Borm}}},
  \bibnamefont{et~al.}, \bibinfo{journal}{arXiv e-prints}
  \bibinfo{eid}{arXiv:2010.03477} (\bibinfo{year}{2020}), \eprint{2010.03477}.

\end{thebibliography}
\end{document}